\documentclass[11pt,a4paper]{article}

\usepackage[utf8]{inputenc}
\usepackage[T1]{fontenc}
\usepackage{lmodern}

\usepackage[a4paper, hmargin=2.6cm, top=2.8cm, bottom=3cm, footskip=1.2cm]{geometry}
\usepackage{microtype}

\usepackage{amsmath}
\usepackage{amssymb}
\usepackage{mathtools}
\usepackage{physics}
\usepackage{booktabs}
\usepackage{mathdots}
\numberwithin{equation}{section}
\allowdisplaybreaks

\usepackage{algorithm}
\usepackage{algpseudocode}

\usepackage{graphicx}
\usepackage{caption}
\usepackage[skip=8pt]{subcaption}

\usepackage[dvipsnames]{xcolor}

\definecolor{LinkBlue}{rgb}{0,0,0.6}
\definecolor{CiteBurgundy}{rgb}{0.4,.2,0.7}
\definecolor{UrlGray}{HTML}{404040}

\usepackage[hyperfootnotes=false, colorlinks=true,
  linkcolor=LinkBlue, citecolor=CiteBurgundy, urlcolor=UrlGray,
  pdftitle={Collective Bootstrap for Matrix Quantum Mechanics},
  pdfauthor={Xianlong Liu}]{hyperref}
\usepackage[noabbrev]{cleveref}

\usepackage{titlesec}
\titleformat{\section}{\Large\bfseries}{\thesection}{0.7em}{}
\titleformat{\subsection}{\large\bfseries}{\thesubsection}{0.7em}{}
\titleformat{\subsubsection}{\normalsize\bfseries}{\thesubsubsection}{0.7em}{}
\titlespacing*{\section}{0pt}{2.2ex plus 0.8ex minus 0.2ex}{1.2ex plus 0.2ex}
\titlespacing*{\subsection}{0pt}{1.8ex plus 0.6ex minus 0.2ex}{0.9ex plus 0.2ex}
\titlespacing*{\subsubsection}{0pt}{1.5ex plus 0.5ex minus 0.2ex}{0.6ex plus 0.1ex}

\usepackage[titles]{tocloft}
\newcommand{\ii}{\mathrm{i}}

\newcommand{\col}{\mathrm{col}}
\newcommand{\len}{\ell}

\definecolor{CommentRed}{rgb}{0.6,0,0}
\definecolor{AdditionBlue}{rgb}{0,0,0.55}
\newif\ifshowedits
\showeditstrue
\ifshowedits
  \newcommand{\comment}[1]{{\color{CommentRed}\sffamily\small[\textbf{Comment:} #1]}}
  
\else
  \newcommand{\comment}[1]{}
  
\fi

\makeatletter
\newcommand{\subtitle}[1]{\gdef\@subtitle{#1}}
\newcommand{\@subtitle}{}
\newcommand{\institution}[1]{\gdef\@institution{#1}}
\newcommand{\@institution}{}
\newcommand{\noteinfo}[1]{\gdef\@noteinfo{#1}}
\newcommand{\@noteinfo}{}
\newcommand{\abstracttext}[1]{\gdef\@abstracttext{#1}}
\newcommand{\@abstracttext}{}
\newcommand{\email}[1]{\gdef\@email{\href{mailto:#1}{\nolinkurl{#1}}}}
\newcommand{\@email}{}

\renewcommand{\maketitle}{%
\begin{titlepage}
  \centering
  \vspace*{\stretch{1}}

  {\LARGE\sffamily \@title \par}
  \ifx\@subtitle\@empty\else
    \vspace{0.6em}
    {\large\sffamily \@subtitle \par}
  \fi

  \vspace{2.2em}
  {\large \@author \par}
  \vspace{0.8em}
  {\small
  \@institution \par}
  \ifx\@email\@empty\else
    \vspace{0.5em}
    {\small\hypersetup{urlcolor=UrlGray}E-mail: \@email\par}
  \fi
  \ifx\@noteinfo\@empty\else
    \vspace{0.8em}
    {\small \@noteinfo \par}
  \fi

  \vspace{3.5em}
  \begin{minipage}{0.86\textwidth}
    {\centering\bfseries Abstract\par}
    \vspace{0.8em}
    {\normalsize\setlength{\parindent}{16pt}\@abstracttext\par}
  \end{minipage}

  \vspace*{\stretch{2}}
  \vspace*{0.5cm}
\end{titlepage}
}
\makeatother

\title{Collective Bootstrap for Matrix Quantum Mechanics}
\author{Xianlong Liu}
\institution{
    Shanghai Center for Complex Physics, School of Physics and Astronomy, \\  
    Shanghai Jiao Tong University, Shanghai 200240, China
}
\email{xianlong_liu@sjtu.edu.cn}
\abstracttext{
    We develop a collective bootstrap framework for large $N$ matrix quantum mechanics formulated directly in the loop space of single trace coordinate moments. Loop joining and splitting operations organize the kinetic constraints systematically and avoid treating momentum containing moments as independent variables. Employing the virial theorem, a Schur complement, and a relaxation of quadratic products of loops, we recast the nonlinear constrained optimization of the collective energy into a semidefinite program, which yields a lower bound on the planar ground state energy at every truncation. A sequential semidefinite optimization heuristic then searches for configurations satisfying planar factorization and estimates the ground state energy and loop expectation values. Applications to one-, two-, and three-matrix models exhibit close agreement with analytic solutions and master field optimization. By working exclusively with coordinate moments, the framework permits truncations at higher word length and offers a promising approach to models with more matrices, including the bosonic sectors of the BFSS and BMN models.
}

\begin{document}
\maketitle

{\hypersetup{linkcolor=black}\tableofcontents}
\thispagestyle{empty}
\newpage
\setcounter{page}{1}

\section{Introduction}
\label{sec:introduction}

Matrix quantum mechanics (MQM) plays a central role in theoretical high energy physics, whose perturbative expansion is organized by planar diagrams \cite{tHooft:1973alw} in the large $N$ limit. The simplest single hermitian MQM is equivalent to free fermions and can be solved exactly \cite{Brezin:1977sv}. In a double-scaling limit this model describes $c=1$ string theory \cite{Klebanov:1991qa,Ginsparg:1993is}, and its collective field formulation provides the corresponding string field theory \cite{Das:1990kaa}. Multi-matrix models arise from the dimensional reduction of supersymmetric Yang--Mills theory. A prominent example is the Banks--Fischler--Shenker--Susskind (BFSS) matrix model \cite{deWit:1988wri,Banks:1996vh}, whose degrees of freedom are nine bosonic Hermitian matrices and their fermionic superpartners. In the large $N$ limit, the model is conjectured to provide a nonperturbative description of M-theory in the infinite-momentum frame \cite{Banks:1996vh,Taylor:2001vb}, and its mass deformation, the BMN model \cite{Berenstein:2002jq}, describes M-theory on a plane-wave background.

Solving for the large $N$ ground state of a generic interacting multi-matrix system remains challenging. Lattice Monte Carlo simulations have provided impressive tests of gauge/gravity duality in the D0-brane matrix model \cite{Anagnostopoulos:2007fw,Catterall:2008yz,Hanada:2013rga,Berkowitz:2016jlq,Pateloudis:2022ijr} and have mapped out the thermal phase structure of its bosonic sector \cite{Bergner:2019rca}. These calculations, however, are naturally formulated at finite $N$ and finite temperature, so that statements about the planar ground state require a two-fold extrapolation to $N \to \infty$ and $T \to 0$, which is computationally expensive and limits the attainable precision. Two complementary numerical approaches developed recently instead work directly at large $N$.

Master field optimization \cite{Jevicki:1983hb,Koch:2021yeb,Mathaba:2023non,Rodrigues:2025sbu,Jevicki:2025ybt} uses the Jevicki--Sakita collective field theory~\cite{Jevicki:1979mb,Jevicki:1980zg} to express the theory in terms of gauge invariant loop variables. In the planar limit the ground state is described by a single classical configuration, i.e., the master field \cite{Gopakumar:1994iq}. Parametrizing the loops by finite matrices then permits a direct unconstrained minimization of the collective potential, so that the infinitely many constraints of the loops are automatically satisfied. The matrix bootstrap \cite{Anderson:2016rcw,Lin:2020mme,Kazakov:2021lel,Han:2020bkb,Berenstein:2021dyf,Cho:2024kxn,Lin:2023owt,Lin:2024vvg,Lin:2025srf}, on the other hand, combines the Schwinger--Dyson (SD) equations, the quantum mechanical analog of the loop equations~\cite{Makeenko:1979pb}, with symmetries and positivity constraints to bound physical observables by semidefinite programming (SDP).

These methods have complementary strengths. The collective master field formulation gives access to both the large $N$ background and its $O(1)$ fluctuation spectrum \cite{Koch:2021yeb,Mathaba:2023non,Rodrigues:2025sbu}, although the cost of the parametrization grows quickly with the number of matrices and with finite length truncation of the loop space. Matrix bootstrap methods can exploit symmetries to reduce the number of independent moments, which has enabled studies of systems with many matrices, including BFSS \cite{Lin:2024vvg,Lin:2025srf}, producing rigorous bounds. In the phase space formulation, however, the operator basis contains both coordinate and momentum matrices, and its rapid growth makes calculations at long word lengths demanding. Moreover, large $N$ factorization renders the SD equations quadratic in the single trace moments. In matrix bootstrap this nonlinearity can be handled, for instance, by a semidefinite relaxation \cite{Kazakov:2021lel,Kazakov:2022xuh}. 

In this work, we develop a collective bootstrap framework that combines elements of these two approaches. The variables are the single trace moments of the coordinate matrices alone. As basic ingredients of the collective field formulation, loop joining and splitting organize the kinetic constraints directly in coordinate loop space, so that momentum containing words never appear as independent variables. The remaining nonlinearities of the collective energy, namely the inverse of the loop joining matrix and the quadratic products of loops in the loop splitting, are removed in two steps. First, combining the virial theorem with a Schur complement replaces the inverse joining matrix by a positive semidefinite block. Second, in the spirit of the relaxation bootstrap \cite{Kazakov:2021lel,Kazakov:2022xuh}, the products of loops are lifted to auxiliary variables. This yields a semidefinite relaxation of the collective energy constrained optimization problem, whose optimum is a lower bound on the planar ground state energy at every truncation. We then employ a sequential semidefinite optimization, based on the log-det heuristic for rank minimization \cite{Fazel2003}, to search for configurations satisfying planar factorization, which provides an estimate of the optimum of the unrelaxed problem.

We apply the method to one-, two-, and three-matrix models. Analytically solvable cases benchmark the effectiveness of the collective bootstrap method. For the one-matrix model the ground state energy agrees with the exact solution to six digits, and for the decoupled two-matrix model the results reproduce the factorization pattern dictated by free independence. The general multi-matrix systems are not analytically solvable, and we compare the results with the existing master field optimization results. For the two- and three-matrix models with Yang--Mills interactions we find close agreement. In the three-matrix case the ground state energy agrees to $0.022\%$, and the dimensionless commutator ratio (``angle'') $\theta \approx 0.7895$ agrees with the master field value $0.791(2)$ \cite{Rodrigues:2025sbu}.

This paper is organized as follows. \Cref{sec:preliminary} reviews the basic ingredients of the phase space matrix bootstrap and the collective formulation of MQM. \Cref{sec:numerical_framework} presents the collective bootstrap and its numerical implementation, and discusses the status of the resulting numbers as bounds and estimates. Applications to one-, two-, and three-matrix models follow in \Cref{sec:example_one-MQM,sec:two-MQM_quartic,sec:two-MQM_Yang-Mills,sec:three-MQM_Yang-Mills}. \Cref{sec:discussion} concludes with possible extensions. The appendices contain a pedagogical $(0+1)$-d vector model example (appendix \ref{app:vector_model}), the analytic bootstrap and SD hierarchy of the one-matrix model (appendix \ref{app:analytic_bootstrap_one_MQM}), and the ray search used to map feasible regions (appendix \ref{app:ray_search}).

\section{Preliminaries}
\label{sec:preliminary}

We consider quantum mechanics of $d$ Hermitian coordinate  matrices $M_a$ and their canonical conjugate momenta $P_a$, with Hamiltonian and canonical commutation relations
\begin{equation}
    H = \frac{1}{2} \sum_{a=1}^{d} \tr(P_a^2) + V(M_1, M_2, \dots, M_d) \, , 
    \quad
    [(M_{a}){}^{i}{}_{j}, (P_{b}){}^{k}{}_{l} ] = \ii \delta_{ab} \delta^{i}{}_{l} \delta^{k}{}_{j} \, .
\end{equation}
The potential $V$ is a gauge invariant function of the coordinate matrices $M_a$, which may contain single or multi trace interactions. We first review the stationarity and positivity conditions used in the phase space matrix bootstrap \cite{Han:2020bkb}, and then introduce the loop variables and the collective Hamiltonian underlying master field optimization \cite{Jevicki:1983hb,Koch:2021yeb}. We then explain the relations between these two formulations, which motivates the collective bootstrap of \cref{sec:numerical_framework}.

\subsection{Phase space bootstrap}

The bootstrap approach \cite{Han:2020bkb} is based on the simple idea of imposing physical conditions, such as stationarity and positivity, by means of the SD equations and symmetries. Let $\langle \mathcal{O} \rangle \equiv \tr(\rho \mathcal{O})$, where $\rho$ is a density matrix that commutes with the Hamiltonian, for instance an energy eigenstate. Then for any operator $\mathcal{O}$ we have the SD equations
\begin{equation} \label{eq:boostrap_SD}
    \langle [H, \mathcal{O}] \rangle = 0 \, .
\end{equation}
In addition, for any collection of operators $\{\mathcal{O}_i\}$, the Gram matrix of expectation values is positive semidefinite,
\begin{equation} \label{eq:bootstrap_Hankel}
    \mathcal{M} \succeq 0 \, , \quad \mathcal{M}_{i j} \equiv \langle \mathcal{O}^{\dagger}_i \mathcal{O}_j \rangle \, .
\end{equation}
In MQM, gauge invariant single trace observables have the form $\tr(\mathcal{C})$, where $\mathcal{C}$ is a word built from the alphabet $\mathsf{A}(M, P; d) \equiv \{I, M_1, \dots, M_d, P_1, \dots, P_d\}$ and $I$ denotes the identity matrix. Since there are infinitely many such words, one truncates the word length at some $L$ and works with the operator basis $\{\tr(\mathcal{C}) \mid \len(\mathcal{C}) \leq L\}$. The SD equations \eqref{eq:boostrap_SD} and the symmetries of the system are then used to eliminate dependent operators, leaving an independent set usually called the search space. Within the search space, the positivity condition \eqref{eq:bootstrap_Hankel} carves out the feasible region. As the truncation $L$ increases, more constraints are included and the feasible region shrinks towards the physical solution. Since the words contain both $M_a$ and their conjugates $P_a$, the number of single trace operators grows rapidly with $L$. Moreover, words that would be related by reordering and cyclicity classically, such as $\tr(M_a P_a M_b)$ and $\tr(M_a M_b P_a)$, may differ by quantum corrections from the canonical commutation relations, so that the operator basis contains a large redundancy whose elimination is cumbersome. Since this formulation involves both the coordinates and their conjugate momenta, we refer to it as the phase space bootstrap.

\subsection{Collective field theory}

Master field optimization, on the other hand, is based on the loop space representation, which organizes the constraints systematically in coordinate loop space. For a word $C$ of length $\len(C)$ built from the alphabet $\{I, M_1, \dots, M_{d} \}$, the collective single trace operators (loops) are defined as
\begin{equation}
    \phi(C) = \frac{1}{N^{\frac{\len(C)}{2} + 1}} \tr(C) \, . 
\end{equation}
These loops do not involve the momentum operators $P_a$ at all. In terms of the loop coordinates $\phi(C)$ and their canonical conjugates $\Pi(C)$, the collective Hamiltonian takes the form \cite{Jevicki:1979mb,Jevicki:1980zg}
\begin{align}
    H_{\col} & = \frac{1}{2 N^2} \sum_{C, C'} \Pi(C) \Omega(C, C') \Pi(C')
    + N^2 V_{\col}(\phi(C)) \, , 
    \\
    V_{\col}(\phi) & = \frac{1}{8} \sum_{C, C'} \omega(C) \Omega^{-1}(C, C') \omega(C')^{\dagger} + V(\phi(C)) \, , 
    \label{eq:V_col}
\end{align}
where $V(\phi)$ denotes the original potential divided by $N^2$ and expressed in terms of the loops. The first term of $V_{\col}$ arises from the Jacobian of the change of variables from matrices to loops. The loop joining matrix $\Omega$ and the loop splitting vector $\omega$ are defined by
\begin{align}
    \Omega(C, C') & \equiv N^{2} \sum_{a=1}^{d} \tr(\pdv{\phi(C)^{\dagger}}{M_a^{\dagger}} \pdv{\phi(C')}{M_a})
    = \sum_{\substack{C'' \\ \len(C'') = \len(C) + \len(C') - 2}} j(C, C'; C'') \phi(C'') \, ,
    \label{eq:loop_joining} \\
    \omega(C) & \equiv \sum_{a=1}^{d} \tr(\pdv[2]{\phi(C)}{M_a^{\dagger}}{M_a})
    = \sum_{\substack{C', C'' \\ \len(C') + \len(C'') = \len(C) - 2}} p(C; C', C'') \phi(C') \phi(C'') \, .
    \label{eq:loop_splitting}
\end{align}
In the large $N$ limit, the collective potential dominates and its minimum gives the ground state energy. Since $\Omega$ is a Gram matrix of gradients, it is positive semidefinite, and the ground state is the solution of the constrained optimization problem in loop space
\begin{align} \label{eq:collective_optimization_problem}
    \text{minimize } & V_{\col}(\phi) \, , \nonumber \\
    \text{subject to } & \Omega(\phi) \succeq 0 \, .
\end{align}
The dependence on $\Omega^{-1}$ makes direct minimization in loop space nonlinear and computationally demanding. Moreover, $\Omega \succeq 0$ is necessary but not sufficient for a finite set of loop values to be realized by an actual matrix distribution. Master field optimization \cite{Koch:2021yeb} addresses both difficulties by parametrizing the loops with finite matrix variables $X_A$, so that the objective function becomes $V_{\col}(\phi(X))$. The joining matrix is then a Gram matrix of gradients with respect to $X_A$, and its positivity is automatic. Standard unconstrained optimization methods, such as conjugate gradients or BFGS, can then be applied to the parametrizing variables.\footnote{
    We denote the master field variables by $X_A$ to distinguish them from the original coordinate matrices $M_a$. The objective function is the planar collective potential, rather than the original finite $N$ quantum Hamiltonian, so that the finite $N$ variational principle does not apply. In practice the loop space is also truncated, which drops a nonnegative contribution to the kinetic term (see \cref{subsec:truncation}). In the limit of a large master field and large loop truncation, master field optimization therefore approaches the planar ground state energy from below.
} 
The stationarity condition is
\begin{equation}
    \pdv{V_{\col}(\phi(X_A))}{X_A} = 0 \, , \quad \forall \, X_A \, ,
\end{equation}
whose equivalent expression in loop space is
\begin{equation} \label{eq:stationary_eq_loop_space}
    \sum_{C'} \Omega(C, C') \pdv{V_{\col}(\phi)}{\phi(C')} = 0 \, , \quad \forall \, C \, .
\end{equation}
Contracting with $Q \equiv \Omega^{-1} \omega$ and using the symmetry of $\Omega$, this condition also implies orthogonality of the loop splitting and the gradient of the collective potential
\begin{equation}
    \sum_{C} \omega(C) \pdv{V_{{\col}}(\phi)}{\phi(C)} = 0 \, .
\end{equation}

\subsection{Relation between the two formulations}

The two formulations contain the same physical ingredients and are closely related. Schematically, the constraints obtained from the SD equations \eqref{eq:boostrap_SD} fall into two classes, depending on whether the word $\mathcal{C}$ contains momentum operators:
\begin{align}
    \text{Kinetic: } & \quad  P_i \notin \mathcal{C} 
    \quad \Rightarrow \quad
    [V, \tr(\mathcal{C})] = 0 \, , 
    \label{eq:kinetic_constraints}
    \\
    \text{Dynamical: } & \quad  P_i \in \mathcal{C}
    \quad \Rightarrow \quad [V, \tr(\mathcal{C})] \neq 0 \, .
    \label{eq:dynamical_constraints}
\end{align}
The joining and splitting operations organize the kinetic relations in loop space, while collective stationarity supplies the corresponding dynamical equations. Appendix~\ref{app_sub:one-MQM_SD_hierarchy} demonstrates this relation in detail for the one-matrix model, and the generic multi-matrix cases follow in a similar manner. The kinetic SD equations determine the momentum containing moments $\langle \tr(\partial \phi(C) / \partial M_a \, P_a) \rangle$ in terms of products of coordinate loops $\phi(C)$. In the collective description these are encoded by $Q \equiv \Omega^{-1} \omega$, and the kinetic SD equations become precisely the statement that $Q$ solves the linear system built from loop joining and splitting. The two sets of relations then take the compact form
\begin{align}
    \text{Kinetic: } & \quad \Omega(C, C') Q(C') = \omega(C) \, , \\
    \text{Dynamical: } & \quad \Omega(C, C') \partial_{C'} V_{\col}(\phi) = 0 \, .
\end{align}
The kinetic relations are thus solved identically once loops are used as variables, and only the dynamical relations remain to be imposed. Collective field theory therefore organizes the constraints in a systematic way.

The collective bootstrap aims to solve the constrained optimization problem \eqref{eq:collective_optimization_problem} directly in coordinate loop space. In the spirit of bootstrap methods, we linearize it at the cost of introducing additional variables and constraints, so that it becomes an SDP. Compared with the phase space bootstrap, this framework has several advantages. Since it works in coordinate space rather than phase space, one never needs to reorder words such as $\tr(MPM)$ and $\tr(MMP)$ using the canonical commutation relations. As a consequence, the number of single trace variables at word length $L$ is reduced from roughly $(2d)^L$ to roughly $d^L$ for a $d$-matrix model, before cyclicity and symmetry reductions. This in turn allows us to explore larger truncations and to determine loops with longer words. Moreover, the method directly maps out feasible regions in the space of coordinate moments. 

\section{Numerical framework}
\label{sec:numerical_framework}

\subsection{The semidefinite programming problem}

We now formulate the collective bootstrap as a semidefinite relaxation followed by a rank minimization. The latter is equivalent to a search for planar factorization $\langle \phi(C) \phi(C') \rangle = \langle \phi(C) \rangle \langle \phi(C') \rangle$. Throughout this paper we denote by $E \equiv \langle H \rangle$ the energy expectation value, by
\begin{equation}
    E_{0} \equiv \min \lim_{N \rightarrow \infty} \frac{1}{N^2} \langle H \rangle
\end{equation}
the exact large $N$ ground state energy divided by $N^2$, and by $E_{\rm{g}}$ its numerical estimate obtained with the collective bootstrap.

The first ingredient is the virial theorem, which states that
\begin{equation}
    2 \langle T \rangle = \sum_{a=1}^{d} \biggl\langle \tr(M_a \pdv{V}{M_a}) \biggr\rangle \, ,
\end{equation}
where $T$ and $V$ denote the kinetic and potential operators, respectively. It follows from substituting $\mathcal{O} = \sum_{a=1}^{d} \tr(M_a P_a)$ into the SD equation \eqref{eq:boostrap_SD}, and is the first nontrivial dynamical constraint in \eqref{eq:dynamical_constraints}. At large $N$ the kinetic energy is $\langle T \rangle / N^2 = \omega^{\rm{T}} \Omega^{-1} \omega^{\dagger} / 8$ in the collective description, so that the virial theorem becomes
\begin{equation} \label{eq:virial}
    \frac{1}{4} \mu 
    = \frac{1}{N^2} \sum_{a=1}^{d} \biggl\langle \tr(M_a \pdv{V}{M_a}(M_1, \dots, M_{d})) \biggr\rangle \, , \quad
    \mu \equiv \omega^{\rm{T}} \Omega^{-1} \omega^{\dagger} \geq 0 \, .
\end{equation}
The right hand side is a linear combination of loops (or products of loops for multi trace potentials). The energy, i.e.\ the on-shell collective potential, therefore becomes
\begin{equation} \label{eq:E_mu_rep}
    \frac{1}{N^2} E =  \frac{1}{8} \mu + V(\phi)
    =  \frac{1}{2 N^2} \sum_{a=1}^{d} \biggl\langle \tr(M_a \pdv{V}{M_a})\biggr\rangle + V(\phi) \, .
\end{equation}

The energy \eqref{eq:E_mu_rep} is linear in the loops, but $\mu$ is still defined through $\Omega^{-1}$. The second ingredient replaces the inverse joining matrix by a positive semidefinite block. Using a Schur complement, we relax the definition of $\mu$ in \cref{eq:virial} to the constraint
\begin{equation} \label{eq:Schur_complement}
    S \equiv \begin{pmatrix}
        \Omega & \omega^{\dagger} \\
        \omega^{\rm{T}} & \mu
    \end{pmatrix} \succeq 0 \, .
\end{equation}
We have
\begin{equation}
    S \succeq 0 
    \quad \Leftrightarrow \quad
    \Omega \succeq 0 \, , \quad \mu - \omega^{\rm{T}} \Omega^{-1} \omega^{\dagger} \geq 0 \, .
\end{equation}
At the physical solution this inequality is saturated, and the definition of $\mu$ in \cref{eq:virial} is recovered. 

The third ingredient linearizes the splitting vector at the cost of introducing auxiliary variables. The elements of the loop joining matrix are linear in $\phi$, whereas the elements of the loop splitting vector $\omega$ \eqref{eq:loop_splitting} are quadratic. To cast the problem as an SDP, we introduce the two-index variables
\begin{equation}
    Y(C, C') \equiv \phi(C)^{\dagger} \phi(C') \, ,
\end{equation}
in terms of which the loop splitting is linear. For the trivial word $C = I$ we have $\phi(I) = \tr(I) / N = 1$, and therefore $Y(I, I) = 1$, $Y(I, C) = \phi(C)$ and $Y(C, I) = \phi(C)^{\dagger}$. The genuinely new variables are $Y(C_1, C_2)$ with both $C_1$ and $C_2$ nontrivial, and the lifted matrix takes the form
\begin{equation}
    Y = \begin{pmatrix}
        1 & \phi(C_1) & \phi(C_2) & \phi(C_3) & \dots \\
        \phi(C_1)^{\dagger} & Y(C_1, C_1) & Y(C_1, C_2) & Y(C_1, C_3) & \dots \\
        \phi(C_2)^{\dagger} & Y(C_2, C_1) & Y(C_2, C_2) & Y(C_2, C_3) & \dots \\
        \phi(C_3)^{\dagger} & Y(C_3, C_1) & Y(C_3, C_2) & Y(C_3, C_3) & \dots \\
        \vdots & \vdots & \vdots & \vdots & \ddots 
    \end{pmatrix} \, , 
    \quad
    C_i \neq I \, .
\end{equation}
Note that $Y$ is the lifted outer product $Y = v^{\dagger} v$ with $v \equiv (1, \phi(C_1), \phi(C_2), \dots)$, i.e.\ a Gram matrix, and hence is positive semidefinite. Moreover, by construction it has rank one for physical solutions. Together these give the consistency conditions
\begin{equation}
    Y \succeq 0 \, , \quad \rank(Y) = 1 \, .
\end{equation}
Dropping the rank condition and keeping only $Y \succeq 0$ is the standard semidefinite relaxation of a quadratically constrained problem \cite{Shor:1987}, which is also the lowest level of the moment hierarchy for polynomial optimization \cite{Lasserre:2001}. 

In the large $N$ matrix bootstrap this lifting was introduced by Kazakov and Zheng \cite{Kazakov:2021lel}. There the loop equations of matrix integrals are quadratic in single trace moments because of large $N$ factorization, and replacing the products by a matrix constrained in this way turns the nonlinear problem into an SDP that yields bounds. The relaxation is not saturated at finite cutoff, but the violation of the quadratic relations was found numerically to decrease as the cutoff is increased. The same relaxation was subsequently applied to lattice Yang--Mills theory \cite{Kazakov:2022xuh}. The present framework uses this relaxation in a different setting. 
The main nonlinearity of the collective problem is not a product of moments but the inverse joining matrix $\Omega^{-1}$ in the kinetic term. It is handled by the virial relation together with the Schur complement \eqref{eq:Schur_complement}, and the lifting is needed only for the quadratic splitting $\omega$. Accordingly, only products of loops of length $\leq \ell_{\max} - 2$ are lifted, so that $Y$ is much smaller than the joining matrix $\Omega$, and no momentum containing words enter.
In addition to the relaxation, which gives the lower bound $E_{\rm{R}}$ discussed in \cref{subsubsec:status_of_results} and plays the role of the relaxed bounds of \cite{Kazakov:2021lel}, we search for rank-one (factorized) points by sequential SDP. This targets the optimum of the unrelaxed problem directly, at the price of using a heuristic. As shown in \cref{sec:example_one-MQM,sec:two-MQM_quartic,sec:two-MQM_Yang-Mills,sec:three-MQM_Yang-Mills}, this search nevertheless yields results in close agreement with analytic solutions and with master field optimization.

With these three ingredients, the problem \eqref{eq:collective_optimization_problem} becomes the following optimization problem. The variables are $\{\phi(C), Y(C, C') \}$, where $Y(C, C')$ with nontrivial $C$ and $C'$ linearize the loop splitting $\omega$. The virial theorem eliminates $\mu$ in favor of loop variables, making the energy \eqref{eq:E_mu_rep} linear. The ground state is then the minimum of the energy subject to the constraints:
\begin{align}
    \text{minimize } & \frac{1}{N^2} E = \frac{1}{8} \mu + V(\phi) 
    \nonumber \\
    \text{subject to } & \begin{pmatrix}
        \Omega(\phi) & \omega^{\dagger}(\phi, Y) \\
        \omega^{\rm{T}}(\phi, Y) & \mu
    \end{pmatrix} \succeq 0 \, , 
    \nonumber  \\
    & \mu = \frac{4}{N^2} 
      \sum_{a=1}^{d} \biggl\langle \tr(M_a \pdv{V}{M_a}(M_1, \dots, M_{d})) \biggr\rangle \, ,
     \nonumber \\
    \text{and } & Y \succeq 0 \, , \quad 
    \rank{(Y)}=1
    \, .
    \label{eq:MQM_collective_bootstrap}
\end{align}

\subsection{Numerical algorithm}

We now turn to the numerical implementation of this problem. Two issues have to be addressed: the truncation of the infinite dimensional loop space, and a strategy for imposing the nonconvex constraint $\rank(Y)=1$.

\subsubsection{Truncation hierarchy}
\label{subsec:truncation}

Let 
\begin{equation}
\mathcal{V}_{\ell} \equiv \operatorname{span}\{\phi(C) \mid \len(C) = \ell \}
\end{equation}
denote the subspace spanned by the independent loops of length $\ell$, so that the loop space is the direct sum $\mathcal{V} = \oplus_{\ell = 0}^{\infty} \mathcal{V}_{\ell}$. Since there are infinitely many loops, numerical studies require a truncation $\ell_{\max}$. As a rule of thumb, $\ell_{\max}$ should at least include the loops appearing in the interactions. Once $\ell_{\max}$ is imposed, the different objects in the problem involve loops of different maximal lengths, as follows. Let 
\begin{equation}
N_{\Omega} \equiv \sum_{\ell=1}^{\ell_{\max}} \dim(\mathcal{V}_{\ell}) 
\end{equation}
denote the size of the truncated loop joining matrix $\Omega$ \cite{Koch:2021yeb}, i.e.\ the number of dynamical degrees of freedom after truncation. The entries of $\Omega$ contain longer loops, of lengths up to $2 \ell_{\max} - 2$, so that there are 
\begin{equation}
N_{\rm{Loops}} \equiv \sum_{\ell = 1}^{2\ell_{\max} - 2} \dim(\mathcal{V}_{\ell})
\end{equation}
loops in total. The loop splitting, on the other hand, only contains loops of lengths up to $\ell_{\max} - 2$. We therefore only need the variables $Y(C, C')$ with $\len(C), \len(C') \leq \ell_{\max} - 2$, and the size of $Y$ is
\begin{equation}
    N_{Y} \equiv 1 + \sum_{\ell = 1}^{\ell_{\max} - 2} \dim(\mathcal{V}_{\ell}) \, ,
\end{equation}
where the additional $1$ accounts for the trivial word $C = I$. We also denote by $N_{\omega} \equiv N_Y - 1$ the number of nontrivial loops appearing in $\omega$. In summary, the variables at truncation $\ell_{\max}$ are
\begin{equation}
    \{\phi(C) , Y(C', C'')
    \mid 
    1 \leq \len(C) \leq 2 \ell_{\max} - 2 , \,
    1 \leq \len(C'), \len(C'') \leq \ell_{\max} - 2\} \, .
\end{equation}

To see how the kinetic constraint depends on the truncation, denote by $\Omega_{\rm{LL}}$ and $\omega_{\rm{L}}$ the blocks involving loops of length $\leq \ell_{\max}$ (low sector, L), and by the subscript H those involving longer loops (high sector). The full Schur complement \eqref{eq:Schur_complement} reads
\begin{equation} \label{eq:Schur_complement_full}
    \begin{pmatrix}
        \Omega_{\rm{LL}} & \Omega_{\rm{LH}} & \omega_{\rm{L}}^{\dagger} \\
        \Omega_{\rm{HL}} & \Omega_{\rm{HH}} & \omega_{\rm{H}}^{\dagger} \\
        \omega_{\rm{L}}^{\rm{T}} & \omega_{\rm{H}}^{\rm{T}} & \mu
    \end{pmatrix} \succeq 0 \, ,
\end{equation} 
giving the constraint
\begin{equation}
    \mu \geq \omega_{\rm{L}}^{\rm{T}} \Omega_{\rm{LL}}^{-1} \omega_{\rm{L}}^{\dagger} + \Delta \, ,
\end{equation}
with
\begin{equation}
    \Delta = v^{\rm{T}} A^{-1} v \geq 0 \, , 
    \quad
    v \equiv \omega_{\rm{H}} - \Omega_{\rm{HL}} \Omega_{\rm{LL}}^{-1} \omega_{\rm{L}} \, ,
    \quad
    A \equiv \Omega_{\rm{HH}} - \Omega_{\rm{HL}} \Omega_{\rm{LL}}^{-1} \Omega_{\rm{LH}} \, .
\end{equation}
Discarding the high sector removes the nonnegative correction $\Delta$ and weakens the constraint, lowering the bound on $\mu$. As the truncation is increased, $\Delta$ is expected to decrease to zero and the precision improves.

\subsubsection{Matrix rank minimization}

\begin{algorithm}[!t]
\caption{Matrix rank minimization via sequential SDP}
\label{alg:matrix_rank_minimization}
\begin{algorithmic}[1]
\Require Initial value $Y_0$, regulator $\varepsilon$, and constraints $f(\phi)$ (optional)
\Require tolerance $\texttt{tol}$, max iteration $\texttt{max\_iter}$
\Ensure Low-rank matrix $Y_k$
\State Initialize $k \gets 0$
\Repeat
    \State Perform SDP optimization \eqref{eq:Y_rank_SDP}:
    \State \qquad $Y_{k+1} = \arg\min_{Y} \tr[(Y_k + \varepsilon I)^{-1} Y]$
    \State \qquad subject to $Y \succeq 0$ and $f(\phi)$
    \State Compute second largest eigenvalue $\lambda_2$ of $Y_{k+1}$
    \State $k \gets k+1$
\Until{$\lambda_2 < \texttt{tol}$ \textbf{or} $k \geq \texttt{max\_iter}$}
\State \Return Optimization results $Y_k$, and second largest eigenvalue $\lambda_2$ of $Y_k$
\end{algorithmic}
\end{algorithm}

We briefly review the log-det heuristic for matrix rank minimization \cite{Fazel2003}. See \cite{Recht:2010rank} for a review of rank minimization and its convex relaxations. Since the rank counts the nonzero eigenvalues, the concave function
\begin{equation} \label{eq:Y_surrogate}
    \tr \log(Y + \varepsilon I) = \sum_i \log(\lambda_i + \varepsilon) \, ,
\end{equation}
serves as a smooth surrogate of $\rank(Y)$, where $\lambda_1 \geq \lambda_2 \geq \dots \geq 0$ are the eigenvalues of $Y$ and $\varepsilon$ is a small positive regulator. Minimizing this function tends to send as many eigenvalues as possible to zero, and thus favors low-rank $Y$. Since the function is concave, it can be minimized iteratively by linearization. Let $Y_{k}$ denote the value of $Y$ at the $k$-th iteration. The Taylor expansion of \eqref{eq:Y_surrogate} around $Y_k$ is 
\begin{equation}
    \tr \log(Y + \varepsilon I) = \tr \log(Y_k + \varepsilon I) + \tr[(Y_k + \varepsilon I)^{-1} (Y - Y_k)] + O((Y - Y_k)^2) \, .
\end{equation}
Dropping constant terms, the objective function for the $(k+1)$-th iteration is $\tr[(Y_k + \varepsilon)^{-1} Y]$. It is linear in $\phi$ and $Y$, so that each iteration is again an SDP, 
\begin{align} 
    \text{minimize } & \tr[(Y_{k} + \varepsilon I)^{-1} Y] \nonumber \\
    \text{subject to } & Y \succeq 0 \, .
    \label{eq:Y_rank_SDP}
\end{align}
To test whether the rank-one condition is reached, we monitor the second largest eigenvalue $\lambda_2$ (in practice the ratio $\lambda_2 / \lambda_1$), which vanishes if $\rank(Y) = 1$. When it falls below a tolerance, we have found a factorized point in the feasible region. The rank minimization algorithm is summarized in \cref{alg:matrix_rank_minimization}. Since the linearization majorizes the concave surrogate, each iteration does not increase $\tr \log(Y + \varepsilon)$, but the iteration may converge to a local minimum.  Failure to find a rank-one point therefore does not by itself exclude the physical solution. Nevertheless, the search performs well for the MQM models studied here, and yields results consistent with analytic solutions and master field calculations.

\subsubsection{Ground state bootstrap: bisection sequential SDP}

\begin{figure}[!t]
\begin{center}
    \includegraphics[width=0.5\textwidth]{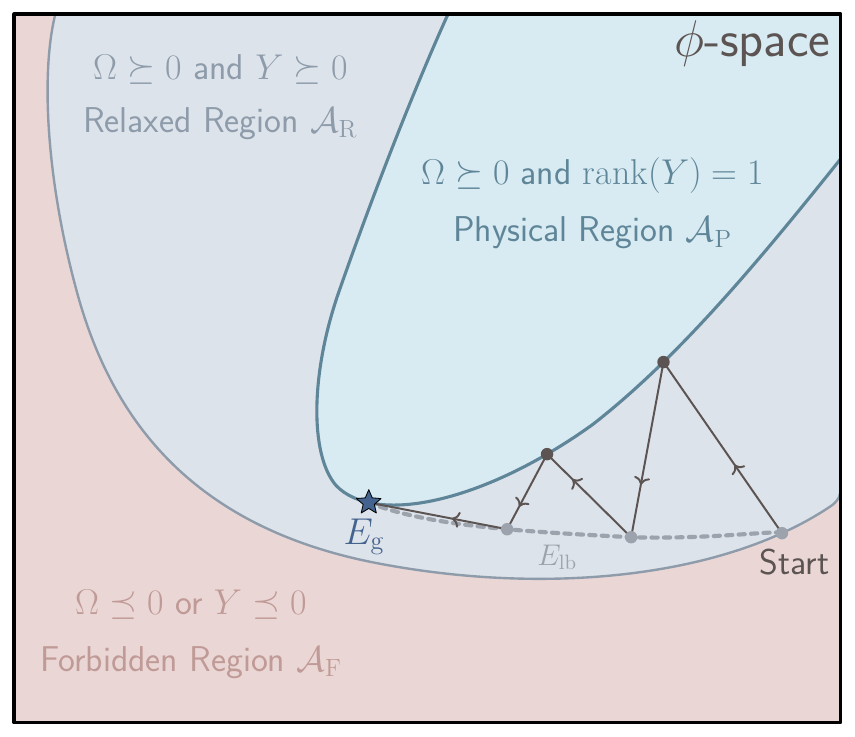}
    \caption{Illustration of the bisection sequential SDP. The relaxed feasible region $\mathcal{A}_{\rm R}$ contains the rank-one region $\mathcal{A}_{\rm P}$. The energy bound $E < E_{\rm test}$ is lowered until no rank-one point is found.} 
    \label{fig:SSDP_illustration}
\end{center}
\end{figure}

\begin{algorithm}[!thb]
\caption{Bisection sequential SDP for large $N$ MQM ground state}
\label{alg:outer_bisection}
\begin{algorithmic}[1]
\Require Coupling parameters, loop space truncation $\ell_{\max}$
\Require Energy bisection tolerance $\varepsilon_E$, rank tolerance $\texttt{tol}$
\Ensure Ground state energy $E_{\mathrm{g}}$, loop expectation values $\langle \phi(C) \rangle$.
\State Eliminate redundant loops by symmetry, and obtain a reduced truncated loop space
\State Perform the SDP \eqref{eq:MQM_collective_bootstrap_outer}, obtain energy lower bound $E_{\rm{lb}}$
\State Set the upper end for the bracket $E_{\rm{ub}}$. For example, a known rank-one point such as a Gaussian state would give a certified initial $E_{\rm ub}$.
\Repeat
    \State $E_{\rm{test}} \gets (E_{\rm{lb}} + E_{\rm{ub}}) / 2$
    \State Perform rank minimization using \cref{alg:matrix_rank_minimization} with additional constraint $E < E_{\rm{test}}$
    \State Obtain the second largest eigenvalue $\lambda_2$ of $Y$
    \If{$\lambda_2 < \texttt{tol}$}
        \State $E_{\rm{ub}} \gets E_{\rm{test}}$
        \Comment{\textit{Rank-one point found: lower $E_{\rm{ub}}$.}}
    \Else
        \State $E_{\rm{lb}} \gets E_{\rm{test}}$
        \Comment{\textit{No rank-one point found: raise $E_{\rm{lb}}$.}}
    \EndIf
\Until{$E_{\rm{ub}} - E_{\rm{lb}} < \varepsilon_{E}$}
\State \Return $E_{\mathrm{g}} = E_{\rm{ub}}$, $\langle\phi(C)\rangle$
\end{algorithmic}
\end{algorithm}

With the truncation and the rank minimization settled, we now describe the algorithm for the large $N$ ground state. Since the rank is not a convex function, we split the problem \eqref{eq:MQM_collective_bootstrap} into two stages. In the first stage we solve the relaxed SDP without imposing $\rank(Y) = 1$:
\begin{align}
    \text{minimize } & \frac{1}{N^2} E = \frac{1}{8} \mu + V(\phi) 
    \nonumber \\
    \text{subject to } & \begin{pmatrix}
        \Omega(\phi) & \omega^{\dagger}(\phi, Y) \\
        \omega^{\rm{T}}(\phi, Y) & \mu
    \end{pmatrix} \succeq 0 \, , 
    \nonumber  \\
    & \mu = \frac{4}{N^2}
      \sum_{a=1}^{d} \biggl\langle \tr(M_a \pdv{V}{M_a}(M_1, \dots, M_{d})) \biggr\rangle \, ,
    \text{ and } Y \succeq 0
    \, .
    \label{eq:MQM_collective_bootstrap_outer}
\end{align}
Denote by $\mathcal{A}_{\rm{R}}$ the feasible region of this relaxed problem, and by $\mathcal{A}_{\rm{P}}$ the feasible region of \eqref{eq:MQM_collective_bootstrap} including $\rank(Y) = 1$, so that $\mathcal{A}_{\rm{P}} \subset \mathcal{A}_{\rm{R}}$. The optimum of the relaxed problem, denoted $E_{\rm R}$ below, provides the initial lower end $E_{\rm lb}$ of the bisection bracket. In the second stage we add the constraint $E < E_{\rm test}$ and run the rank minimization of \cref{alg:matrix_rank_minimization} to decide whether $\mathcal{A}_{\rm{P}}$ contains a point below $E_{\rm test}$. Bisecting in $E_{\rm test}$ locates the lowest energy in $\mathcal{A}_{\rm{P}}$. The procedure is illustrated in \cref{fig:SSDP_illustration} and summarized in \cref{alg:outer_bisection}. A minimal example for the $(0+1)$-dimensional vector model is worked out in appendix \ref{app:vector_model}.

\subsubsection{Status of the numerical results}
\label{subsubsec:status_of_results}

It is useful to state precisely what the quantities computed above really represent. Let $E_0$ denote the exact large $N$ ground state energy (divided by $N^2$), with a loop configuration $\phi^{*}$ that is realized by some matrix distribution. Then $\{ \phi_i^{*}, Y_{ij}^{*} \equiv \bar{\phi}_i^{*} \phi_j^{*} \}$ is a feasible point of the truncated problem \eqref{eq:MQM_collective_bootstrap} for every $\ell_{\max}$, at which the objective function equals $E_0$. The truncations described in this section enlarge the feasible set, including dropping the high sector $\Delta \geq 0$ in \eqref{eq:Schur_complement_full} and restricting $Y$ to loops of length $\leq \ell_{\max} - 2$. Denoting by $E_{\rm{R}}(\ell_{\max})$ and $E_{\rm{P}}(\ell_{\max})$ the exact optima of the relaxed problem \eqref{eq:MQM_collective_bootstrap_outer} and of the rank-constrained problem \eqref{eq:MQM_collective_bootstrap} at truncation $\ell_{\max}$, we therefore have
\begin{equation} \label{eq:bound_hierarchy}
    E_{\rm{R}}(\ell_{\max}) \leq E_{\rm{P}}(\ell_{\max}) \leq E_0 \, .
\end{equation}
Since the constraints at truncation $\ell_{\max}$ contain those at any smaller truncation as principal submatrices, both $E_{\rm{R}}(\ell_{\max})$ and $E_{\rm{P}}(\ell_{\max})$ are nondecreasing in $\ell_{\max}$. The first stage computation of $E_{\rm{R}}(\ell_{\max})$ \cref{alg:outer_bisection} therefore provides a lower bound on $E_0$ up to solver tolerance.

The rank-constrained optimum $E_{\rm{P}}$ is not computed rigorously. If the rank minimization of \cref{alg:matrix_rank_minimization} always found a rank-one point whenever one exists, the bisection would converge to $E_{\rm{P}}(\ell_{\max})$ within some tolerance $\varepsilon_E$. However, \cref{alg:matrix_rank_minimization} is a local heuristic: it may fail at a test energy $E_{\rm{test}} > E_{\rm{P}}$ at which rank-one points exist, and the bisection then moves the bracket upward. On the other hand, every accepted test energy is certified by an approximate rank-one point, so that the returned value satisfies $E_{\rm{g}} \geq E_{\rm{P}}(\ell_{\max})$ up to the tolerances $\varepsilon_E$ and $\texttt{tol}$. The rigorous statements are
\begin{equation}
    E_{\rm{R}}(\ell_{\max}) \leq E_{\rm{P}}(\ell_{\max}) \leq \min\bigl(E_0, E_{\rm{g}}\bigr) \, ,
\end{equation}
while $E_{\rm{g}}$ and $E_0$ are not ordered a priori. As $\ell_{\max}$ grows, $E_{\rm{P}}(\ell_{\max})$ approaches $E_0$ from below; a value $E_{\rm{g}} > E_0$ therefore signals that the rank-one search stopped above the optimum of the truncated problem, and not a violation of \eqref{eq:bound_hierarchy}. We emphasize that \eqref{eq:bound_hierarchy} relies only on constraints satisfied by the exact solution. Regarding loop expectation values, since the high lying loops of length close to $2 \ell_{\max} - 2$ are less constrained, they are therefore determined less precisely than the low lying loops of short lengths.

Despite these caveats, the method is effective in practice. In all models studied below, the energies and the low lying loops agree with analytic solutions and with master field optimization, with relative deviations in the energy ranging from below $10^{-6}$ in the one-matrix model to a few times $10^{-3}$ in the Yang--Mills models. The gap between the bound $E_{\rm{R}}$ and the estimate $E_{\rm{g}}$ is due to the rank-one search alone, and improving this step, e.g.\ by alternative rank-reduction schemes or by global methods for the nonconvex constraint, is left for future work.

\section{One-matrix model}
\label{sec:example_one-MQM}

\begin{figure}[!t]
\begin{center}
    \includegraphics[width=0.5\textwidth]{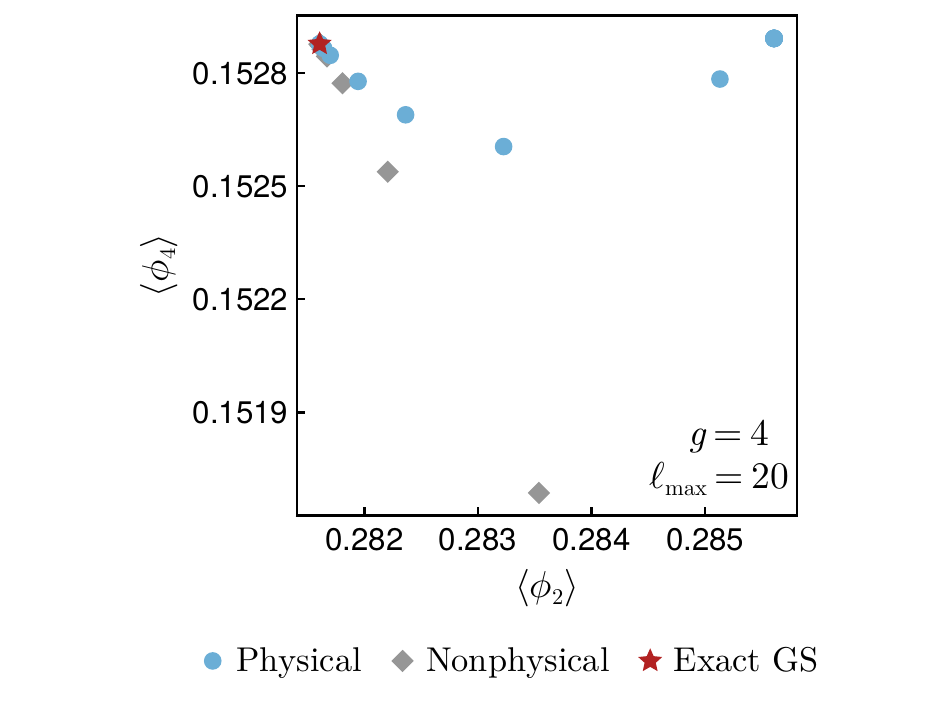}
    \caption{Optimization history in the $(\phi_2, \phi_4)$ plane of the bisection sequential SDP for the one-matrix model at $g = 4$ with truncation $\ell_{\max}=20$.} 
    \label{fig:one-MQM_history}
\end{center}
\end{figure}

\begin{figure}[!t]
\begin{center}
    \includegraphics[width=0.99\textwidth]{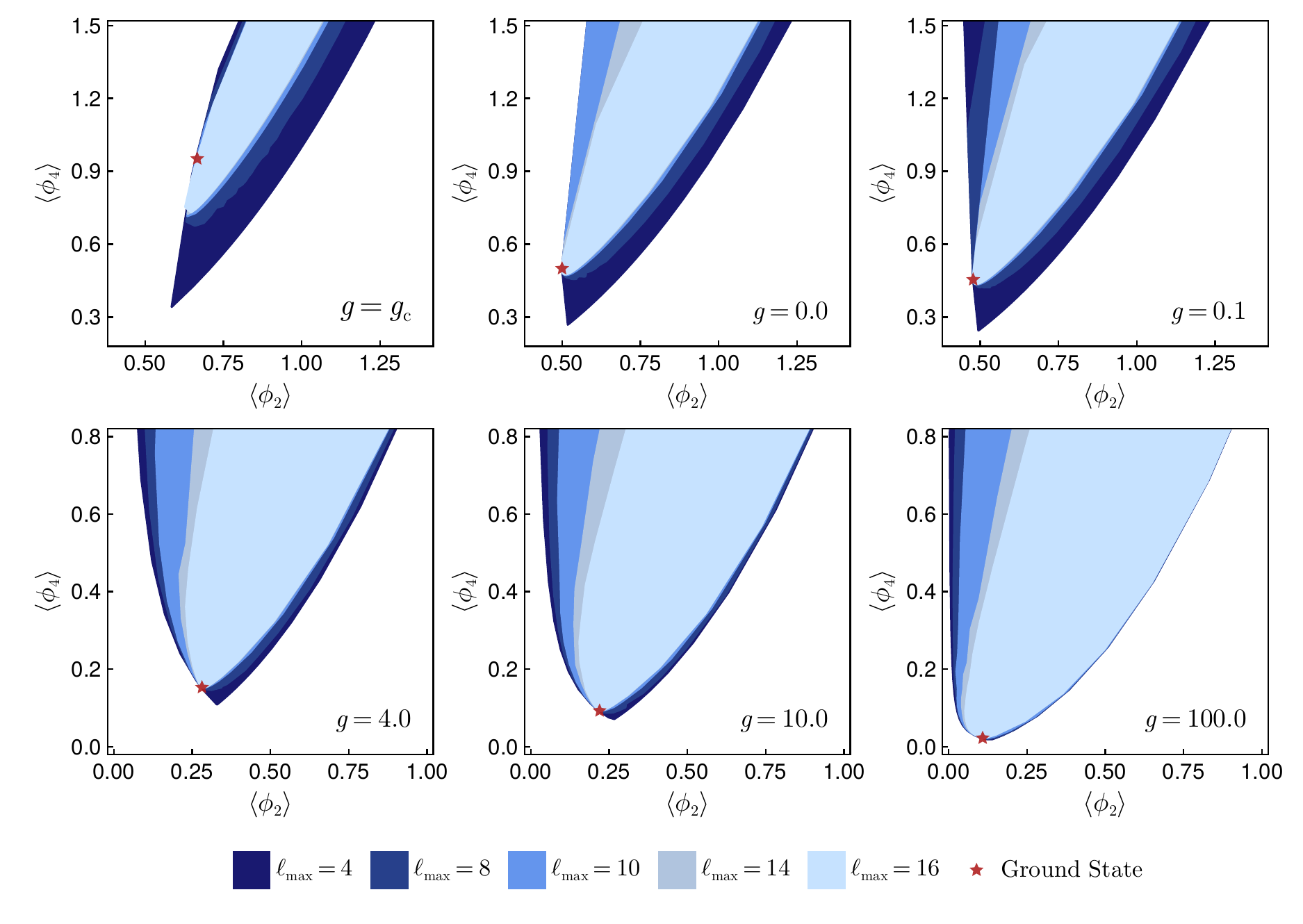}
    \caption{Feasible regions in the projected $(\phi_2, \phi_4)$ subspace of the one-matrix model for various $g$ and different truncations, obtained with the ray search of Appendix~\ref{app:ray_search}. The exact ground states are marked in red stars.}
    \label{fig:one-MQM_feasible_regions}
\end{center}
\end{figure}

We first consider Hermitian one-matrix quantum mechanics with a quartic potential,
\begin{equation} \label{eq:one-MQM_Hamiltonian}
    H = \tr(\frac{1}{2} P^2 + \frac{1}{2} M^2 + \frac{g}{4 N} M^4) \ , 
    \quad
    [M^{i}{}_{j}, P^{k}{}_{l}] = \ii \, \delta^{i}{}_{l} \, \delta^{k}{}_{j} \, .
\end{equation}
This model serves as an ideal benchmark, since it is exactly solvable at large $N$ by mapping to free fermions \cite{Brezin:1977sv}. At large $N$ a stable planar ground state also exists for negative $g$, down to the critical value $g_{\rm{c}} = - 4/(3 \sqrt{2} \pi)$ at which the Fermi level reaches the top of the potential. The loop variables are
\begin{equation}
    \phi_{n} = \frac{1}{N^{\frac{n}{2}+1}} \tr(M^n) \, , \quad n = 0, 1, 2, \dots
\end{equation}
The loop joining matrix and loop splitting vector are
\begin{align}
    \Omega_{mn} & = m n \phi_{m + n - 2} \, , 
    \label{eq:one-MQM_loop_joining} \\
    \omega_{n} & = n \sum_{k=0}^{n-2} \phi_{k} \phi_{n-k-2} \, ,
    \label{eq:one-MQM_loop_splitting} 
\end{align}
and the collective Hamiltonian is
\begin{align}
    H_{\col} & = \frac{1}{2 N^2} \Pi_{m} \Omega_{mn} \Pi_{n} + N^2 V_{\col}(\phi) \, , \\
    V_{\col}(\phi) & = \frac{1}{8} \omega_m (\Omega^{-1})_{m n} \omega_{n} + \frac{1}{2} \phi_2 + \frac{g}{4} \phi_4 \, .
\end{align}
With $\mu \equiv \omega^{\rm{T}} \Omega^{-1} \omega$, the virial theorem gives $ \mu = 4(\phi_2 + g \phi_4)$, and hence the energy is
\begin{equation} \label{eq:E_oneMQM}
    \frac{1}{N^2} E =  \frac{1}{8} \mu + \frac{1}{2} \phi_2 + \frac{g}{4} \phi_4
    = \phi_2 + \frac{3 g}{4} \phi_4 \, .
\end{equation}
Imposing the positivity constraints and the rank-one condition defines the feasible region. \Cref{fig:one-MQM_feasible_regions} shows its projection onto the $(\phi_2, \phi_4)$ plane for several couplings and truncations. Since the energy \eqref{eq:E_oneMQM} is linear in the loops, the ground state lies on the boundary of the feasible region, and as the truncation is increased more constraints are included and the region shrinks. Note that in one-matrix models $\Omega_{mn} = m n \phi_{m+n-2}$ is the Hankel matrix of the moments up to the diagonal rescaling by $m n$, so that $\Omega \succeq 0$ coincides with the usual moment positivity $\mathsf{H} \succeq 0$ with $\mathsf{H}_{ij} \equiv \langle \phi_{i+j} \rangle$. For comparison, the analytic solution is \cite{Brezin:1977sv}
\begin{equation}
    \phi_{n} = \frac{1 + (-1)^n}{2} \frac{\Lambda^{n+2} \sqrt{1 + g \Lambda^2 / 2}}{2 \sqrt{\pi}} \frac{\Gamma(\frac{n}{2} + \frac{1}{2})}{\Gamma(\frac{n}{2} + 2)} {}_2 F_{1}\Bigl(-\frac{1}{2}, \frac{n}{2} + \frac{1}{2} ; \frac{n}{2} + 2 ; z \Bigr) \, ,
\end{equation}
where
\begin{equation}
    z \equiv - \frac{g \Lambda^2 / 2}{1 + g \Lambda^2 / 2} \, , 
    \quad
    \phi_0 = \int_{-\Lambda}^{\Lambda} \rho(x) \dd{x} = 1 \, .
\end{equation}
The normalization condition $\phi_0 = 1$ determines $\Lambda$.
\begin{table}[!t]
\renewcommand{\arraystretch}{1.1}
\begin{center}
\caption{Ground state energy and loop expectation values of the one-matrix model at $g=4$ for several truncations using algorithm \ref{alg:outer_bisection}, compared with the analytic solution. The SDPs are solved with \texttt{SemidefiniteOptimization} in \texttt{Mathematica}, with options \texttt{Method -> "MOSEK"} \cite{MOSEK}, \texttt{PerformanceGoal -> "Quality"}, and \texttt{Tolerance -> 10\textasciicircum(-10)}. Odd loops such as $\phi_1$ and $\phi_3$ vanish by the $\mathbb{Z}_2$ symmetry $M \to -M$.}
\label{tab:one-MQM_bootstrap}
\begin{tabular}{c @{\hspace{3em}} c @{\hspace{1.5em}} c @{\hspace{1.5em}} c @{\hspace{1.5em}} c @{\hspace{1.5em}} c @{\hspace{3em}} c}
    \toprule
    $\ell_{\max}$ & 8 & 12 & 16 & 20 & 24 & analytic \\
    \midrule
    $E_{\rm{g}}$ & 0.740241 & 0.740235 & 0.740235 & 0.740235 & 0.740235 & 0.740235 \\
    \midrule
    $\phi_{2}$  & 0.281619 & 0.281609 & 0.281628 & 0.281615 & 0.281639 & 0.281606 \\
    $\phi_{4}$  & 0.152874 & 0.152875 & 0.152869 & 0.152873 & 0.152866 & 0.152876 \\
    $\phi_{6}$  & 0.102300 & 0.102355 & 0.102330 & 0.102348 & 0.102314 & 0.102361 \\
    $\phi_{8}$  & 0.076286 & 0.076271 & 0.076228 & 0.076256 & 0.076200 & 0.076278 \\
    $\phi_{10}$ & 0.060315 & 0.060675 & 0.060624 & 0.060660 & 0.060587 & 0.060689 \\
    $\phi_{12}$ & 0.049083 & 0.050473 & 0.050400 & 0.050443 & 0.050355 & 0.050477 \\
    $\phi_{14}$ & 0.040484 & 0.043291 & 0.043268 & 0.043316 & 0.043214 & 0.043355 \\
    $\phi_{16}$ &          & 0.037893 & 0.038056 & 0.038113 & 0.037999 & 0.038156 \\
    $\phi_{18}$ &          & 0.033604 & 0.034096 & 0.034181 & 0.034056 & 0.034229 \\
    $\phi_{20}$ &          & 0.030049 & 0.030980 & 0.031130 & 0.030995 & 0.031182 \\
    $\phi_{22}$ &          & 0.027827 & 0.028450 & 0.028709 & 0.028568 & 0.028769 \\
    $\phi_{24}$ &          &          & 0.026332 & 0.026749 & 0.026609 & 0.026827 \\
    $\phi_{26}$ &          &          & 0.024514 & 0.025133 & 0.025006 & 0.025242 \\
    $\phi_{28}$ &          &          & 0.022917 & 0.023777 & 0.023676 & 0.023937 \\
    $\phi_{30}$ &          &          & 0.022413 & 0.022619 & 0.022558 & 0.022853 \\
    $\phi_{32}$ &          &          &          & 0.021614 & 0.021610 & 0.021948 \\
    $\phi_{34}$ &          &          &          & 0.020726 & 0.020795 & 0.021190 \\
    $\phi_{36}$ &          &          &          & 0.019932 & 0.020086 & 0.020553 \\
    $\phi_{38}$ &          &          &          & 0.019533 & 0.019464 & 0.020020 \\
    $\phi_{40}$ &          &          &          &          & 0.018916 & 0.019575 \\
    $\phi_{42}$ &          &          &          &          & 0.018444 & 0.019205 \\
    $\phi_{44}$ &          &          &          &          & 0.018091 & 0.018901 \\
    $\phi_{46}$ &          &          &          &          & 0.018167 & 0.018656 \\
    \bottomrule
\end{tabular}
\end{center}
\end{table}

\begin{figure}[!t]
\begin{center}
    \includegraphics[width=0.5\textwidth]{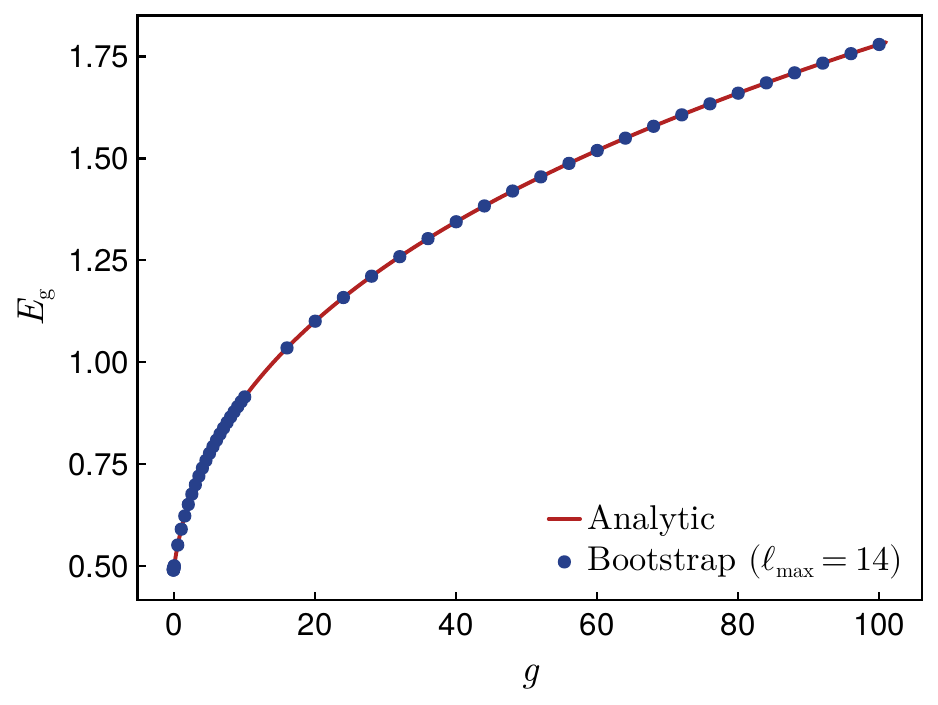}
        \caption{Ground state energy of the one-matrix model versus the coupling $g$ at $\ell_{\max} = 14$, compared with the analytic solution.}
    \label{fig:1M_E_vs_g}
\end{center}
\end{figure}

\begin{figure}[!h]
\begin{center}
    \begin{subfigure}[t]{0.45\textwidth}
        \includegraphics[width=1\textwidth]{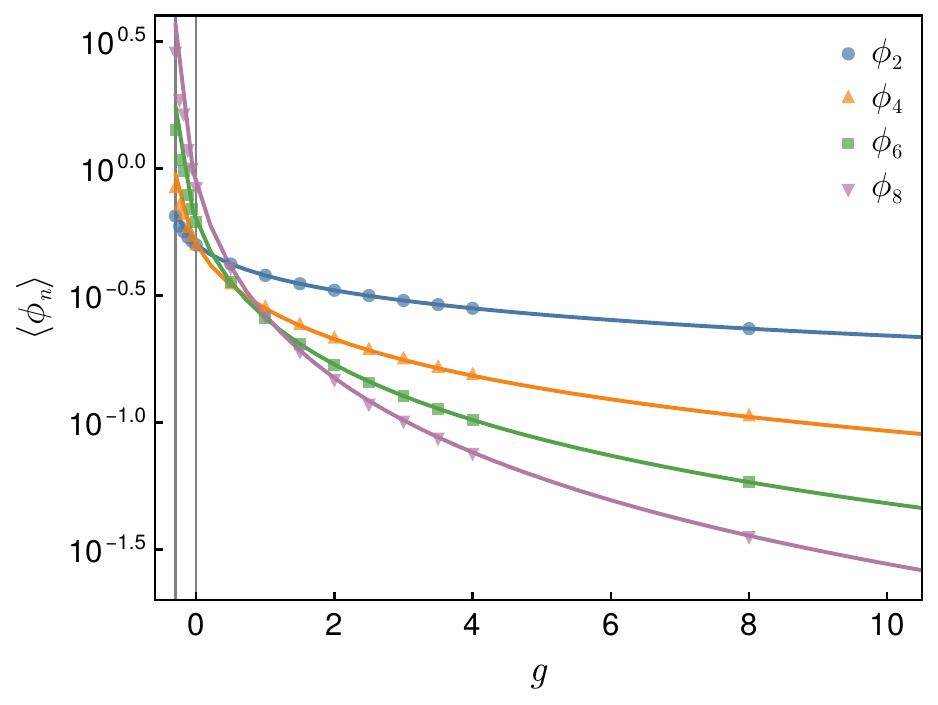}
    \end{subfigure}
    \begin{subfigure}[t]{0.45\textwidth}
        \includegraphics[width=1\textwidth]{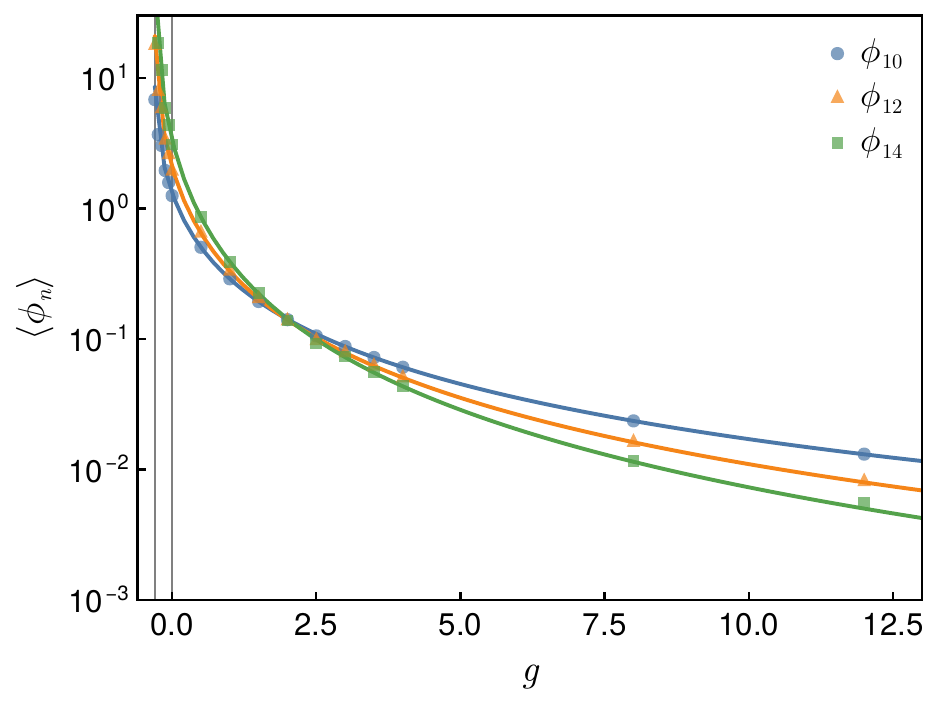}
    \end{subfigure}
    \caption{Loop expectation values of the one-matrix model versus the coupling $g$ at $\ell_{\max} = 14$. Markers are bootstrap results and lines are the exact solution.}
    \label{fig:1M_loops_vs_g}
\end{center}
\end{figure}

We now present the large $N$ ground state obtained with the bisection sequential SDP. As a first test, we fix $g=4$ and vary the truncation. \Cref{tab:one-MQM_bootstrap} lists the ground state energy $E_{\rm{g}}$ and the loop values, with the analytic results in the last column. A typical optimization history of the bisection sequential SDP is shown in \cref{fig:one-MQM_history}. For $\ell_{\max} \geq 12$ the ground state energy agrees with the analytic value $E_0 = 0.7402347$ to all six digits shown. At $\ell_{\max} = 8$ we find $E_{\rm{g}} = 0.740241$, which lies above $E_0$ by $6 \times 10^{-6}$. As explained in \cref{subsubsec:status_of_results}, the exact optimum of the truncated problem cannot exceed $E_0$, so this small excess reflects the rank-one search terminating slightly above the truncated optimum rather than a deficiency of the relaxation. The loop values are less precise than the energy. The loops $\phi_2$ and $\phi_4$ agree with the analytic results to a relative precision of about $10^{-4}$ at all truncations. The longest loops, on the other hand, at each truncation deviate by up to a few percent (e.g.\ $\phi_{46}$ at $\ell_{\max} = 24$ differs from the analytic value by $2.6\%$). This is expected since they are less constrained.
Next, we fix $\ell_{\max} = 14$ and vary the coupling $g$. \Cref{fig:1M_E_vs_g,fig:1M_loops_vs_g} compare the ground state energy and the loops with the analytic results, and show agreement over the whole range of couplings, including the approach to the critical coupling $g_{\rm c}$.

\section{Two-matrix model with quartic interactions}
\label{sec:two-MQM_quartic}

We now turn to multi-matrix models, starting with the two-matrix model with quartic interactions studied in \cite{Koch:2021yeb},
\begin{equation}
    H = \tr(\frac{1}{2} P_1^2 + \frac{1}{2} P_2^2 + \frac{1}{2} M_1^2 + \frac{1}{2} M_2^2 + k M_1 M_2 + \frac{g}{4 N} M_1^4 + \frac{g}{4 N} M_2^4) \, .
\end{equation}
We first consider the exactly solvable case $k = 0$, which provides a nontrivial test of the multi-matrix machinery, and then the generic case $k \neq 0$.

\subsection{The decoupled case: \texorpdfstring{$k=0$}{k=0}}

In this case the model has the discrete symmetry group
\begin{equation}
    \mathsf{G}_1 = \langle \sigma_1, \sigma_2, \tau \rangle
\end{equation}
where
\begin{equation} \label{eq:2MQM_k0_symmetry}
    \sigma_1(M_1, M_2) = (-M_1, M_2) \, , 
    \quad
    \sigma_2(M_1, M_2) = (M_1, -M_2) \, , 
    \quad
    \tau(M_1, M_2) = (M_2, M_1) \, .
\end{equation}
Loops related by these transformations have the same expectation value, $\langle \phi(\mathsf{g}(C)) \rangle = \langle \phi(C) \rangle$ for all $\mathsf{g} \in \mathsf{G}_1$. As a result, only loops with even numbers of both $M_1$ and $M_2$ can have nonzero expectation values, and loops related by $M_1 \leftrightarrow M_2$ have equal expectation values. For example, $\langle \phi(M_1 M_2) \rangle = \langle \phi(M_1 M_1 M_2) \rangle = 0$, and $\langle \phi(M_1 M_1) \rangle = \langle \phi(M_2 M_2) \rangle$.

\begin{figure}[!t]
\begin{center}
    \includegraphics[width=0.95\textwidth]{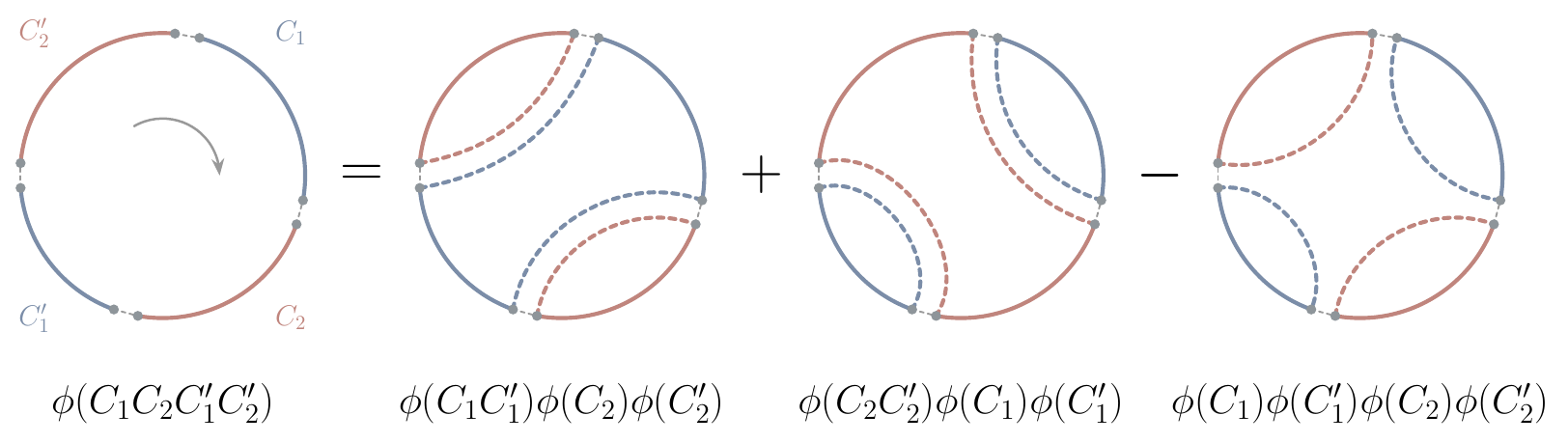}
    \caption{Diagrammatic illustration of the factorization rule \eqref{eq:2MQM_k0_factorization_l4} for the decoupled two-matrix model ($k=0$).}
    \label{fig:two-MQM_factorization_illustration}
\end{center}
\end{figure}

Since the two matrices decouple, the model can be solved using the one-matrix results. The ground state energy is twice the one-matrix energy at the same coupling $g$. Mixed loops are determined at $N = \infty$ by a factorization rule: if $C_1$ involves only $M_1$ and $C_2$ involves only $M_2$, then
\begin{equation}
    \langle \phi(C_1 C_2) \rangle = \langle \phi(C_1) \rangle \langle \phi(C_2) \rangle \, .
\end{equation}
If $C_1$ and $C_1'$ involve only $M_1$, and $C_2$ and $C_2'$ involve only $M_2$, then
\begin{equation} \label{eq:2MQM_k0_factorization_l4}
\begin{split}
    \langle \phi(C_1 C_2 C_1' C_2') \rangle = {} &
    \langle \phi(C_1 C_1') \rangle \langle \phi(C_2) \rangle \langle \phi(C_2') \rangle
    + \langle \phi(C_1) \rangle \langle \phi(C_1') \rangle \langle \phi(C_2 C_2') \rangle
    \\
    & - \langle \phi(C_1) \rangle \langle \phi(C_1') \rangle \langle \phi(C_2) \rangle \langle \phi(C_2') \rangle \, ,
\end{split}
\end{equation}
and so on. More generally, let $C_1, C_1', C_1'', \dots$ denote words involving only $M_1$, and $C_2, C_2', C_2'', \dots$ words involving only $M_2$. Defining the centered words $C^{*} \equiv C - \phi(C) I$, all alternating products of centered words vanish,
\begin{equation}
    \langle \phi(C_1^{*} C_2^{*} C_1^{\prime *} C_2^{\prime *} C_1^{\prime\prime *} C_2^{\prime\prime *} \dots) \rangle = 0 \, .
\end{equation}
This is the statement that $M_1$ and $M_2$ are freely independent in the large $N$ limit \cite{Voiculescu:1992free,Gopakumar:1994iq}, and eq.~\eqref{eq:2MQM_k0_factorization_l4} is the corresponding free moment formula. The factorization rule is illustrated in \cref{fig:two-MQM_factorization_illustration}.

We perform the collective bootstrap at truncation $\ell_{\max} = 8$, for which $N_{\Omega} = 93$ and $N_{\rm{Loops}} = 2615$. Using the symmetry \eqref{eq:2MQM_k0_symmetry}, only 294 independent loops can be nonzero. Since there are 8 independent nonzero loops of length $\len \leq \ell_{\max} - 2 = 6$, there are $8 \times 9 / 2 = 36$ lifted variables $Y$, and altogether 330 variables at $\ell_{\max} = 8$. For comparison, master field optimization requires at least $51^2 + 51 = 2652$ real variables for parametrization of the two matrices at the truncation $\ell_{\max}=8$.

\begin{figure}[!t]
\begin{center}
    \includegraphics[width=0.45\textwidth]{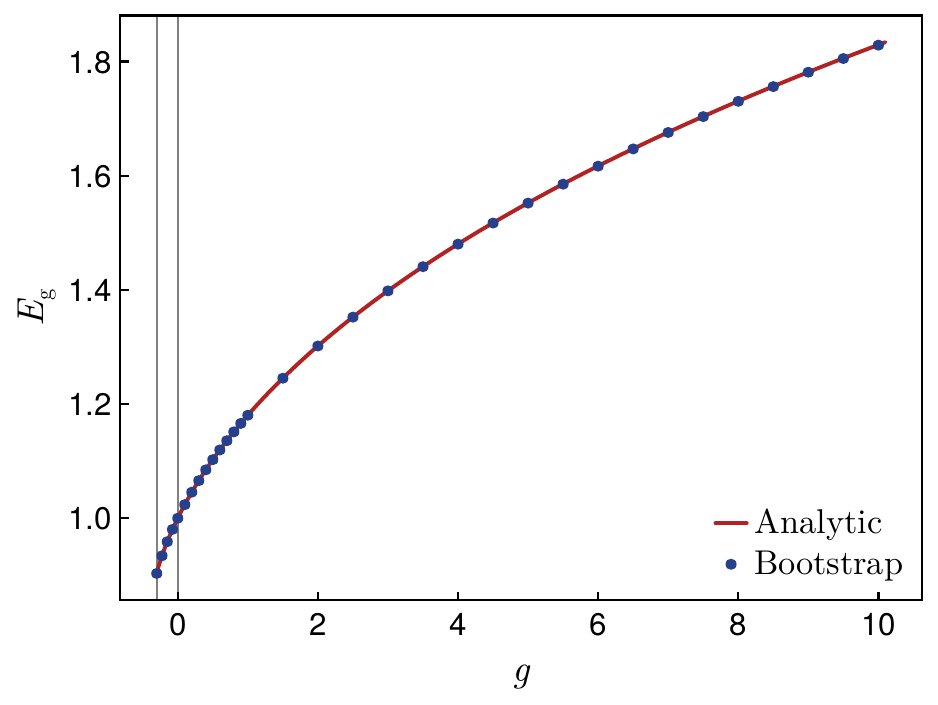}
    \caption{Ground state energy of the decoupled two-matrix model ($k = 0$) versus $g$ at truncation $\ell_{\max} = 8$.}
    \label{fig:2MQM_Eg_vs_g_k0}
\end{center}
\end{figure}

\begin{figure}[!hbt]
\begin{center}
    \includegraphics[width=0.45\textwidth]{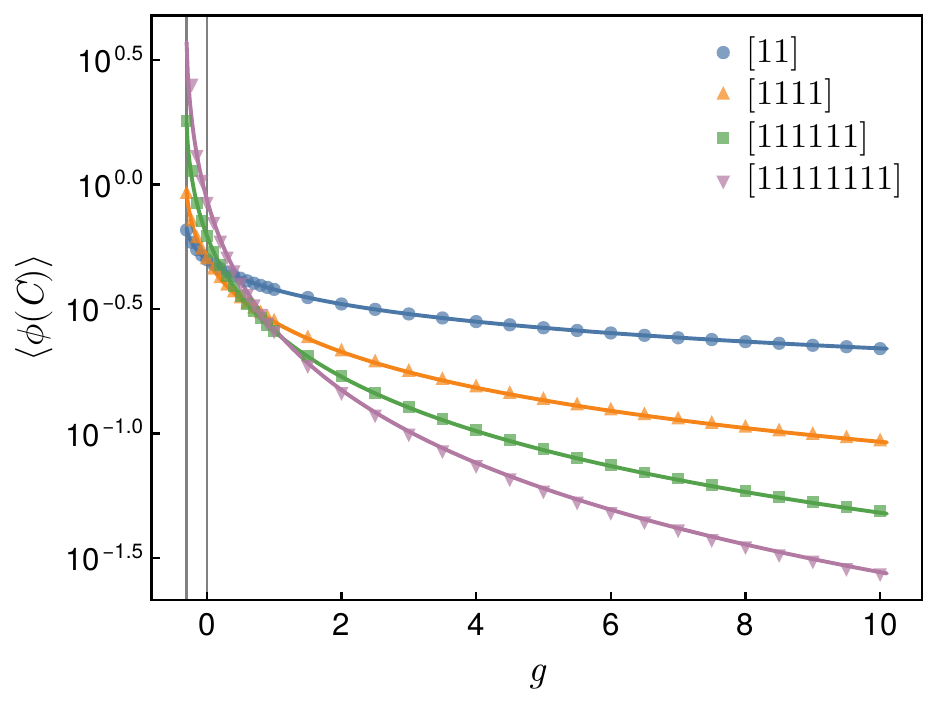}
    \caption{Loops $\phi(M_1^n)$ of the decoupled two-matrix model ($k=0$) at truncation $\ell_{\max} = 8$. Markers are bootstrap results and lines are the exact solution.}
    \label{fig:2MQM_loop_k0_set1}
\end{center}
\end{figure}

\begin{figure}[!t]
\begin{center}
    \begin{subfigure}[t]{0.45\textwidth}
        \includegraphics[width=1\textwidth]{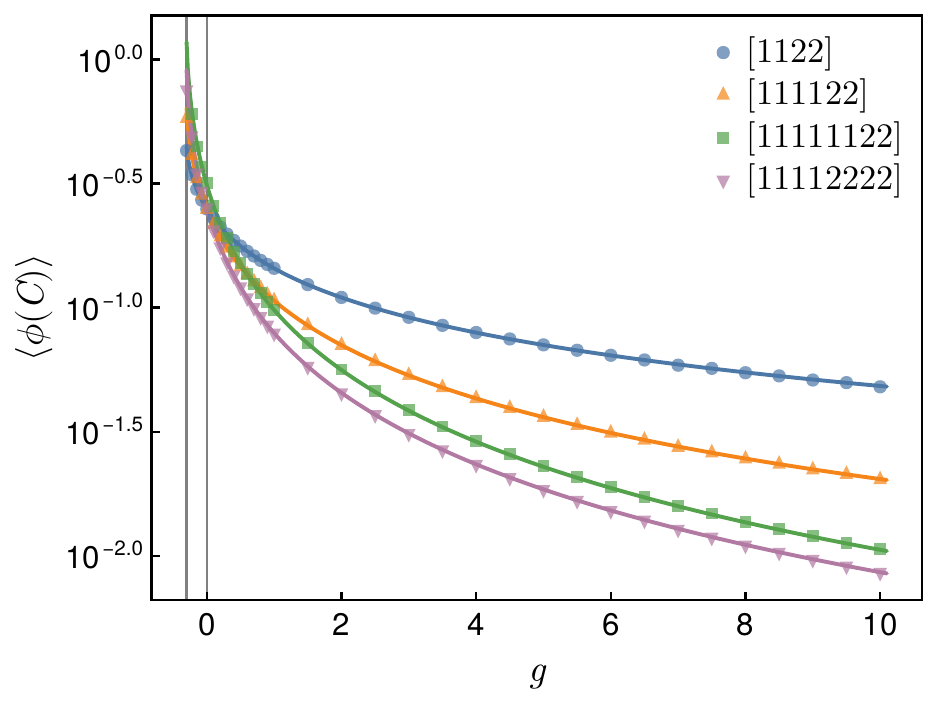}
    \end{subfigure}
    \begin{subfigure}[t]{0.45\textwidth}
        \includegraphics[width=1\textwidth]{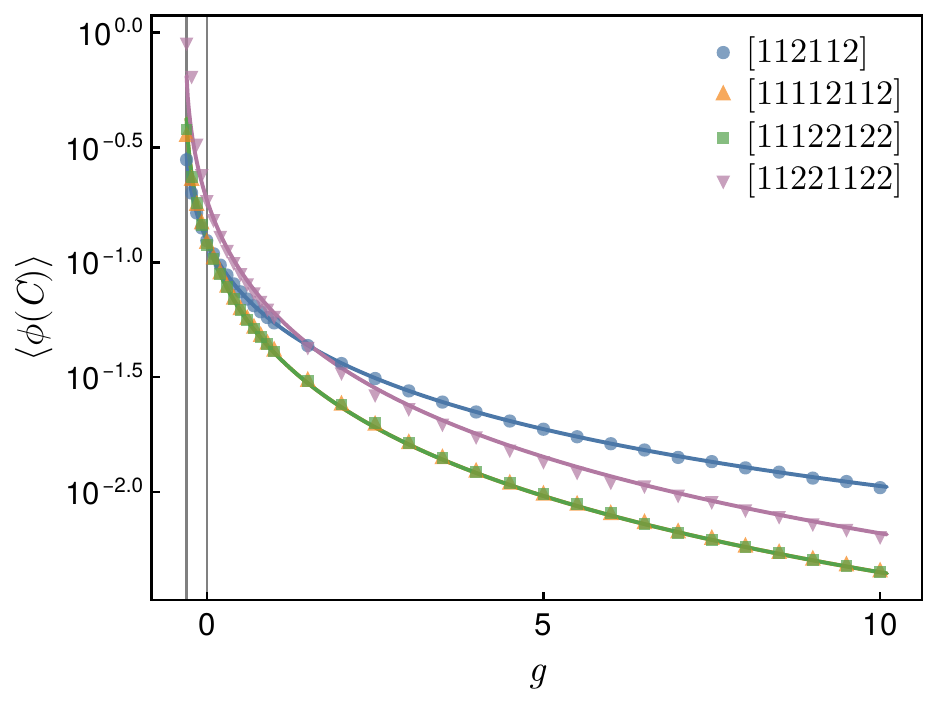}
    \end{subfigure}
    \caption{Mixed loops of the types $\phi(C_1 C_2)$ (left) and $\phi(C_1 C_2 C_1' C_2')$ (right) for the decoupled two-matrix model ($k=0$) at $\ell_{\max} = 8$. Markers are bootstrap results and lines are the exact solution.}
    \label{fig:2MQM_loop_k0_set2_and_3}
\end{center}
\end{figure}

\Cref{fig:2MQM_Eg_vs_g_k0} shows the ground state energy obtained with \texttt{MOSEK}, which agrees with twice the one-matrix result. We use the shorthand notation $[12]$ for $\phi(M_1 M_2)$, and so on. \Cref{fig:2MQM_loop_k0_set1} shows the loops $\phi(M_1^n)$, which should coincide with the one-matrix results at the same coupling; the bootstrap data (markers) agree with the exact results (lines of the same color) over the whole range of couplings. \Cref{fig:2MQM_loop_k0_set2_and_3} shows mixed loops of the types $\phi(C_1 C_2)$ and $\phi(C_1 C_2 C_1' C_2')$, which again agree with the factorization rule. For instance, $[11112112] = \phi(M_1^4 M_2 M_1^2 M_2)$ and $[11122122] = \phi(M_1^3 M_2^2 M_1 M_2^2)$ both equal $\phi_4 \phi_2^2$ by \eqref{eq:2MQM_k0_factorization_l4}, and indeed coincide in the figure. Moreover, loops such as $\phi(M_1 M_2 M_1 M_2)$ are allowed by the symmetry \eqref{eq:2MQM_k0_symmetry} but vanish by free independence. The bootstrap reproduces this and we therefore do not show them. 

\subsection{The generic case: \texorpdfstring{$k\neq0$}{k!=0}}

For $k \neq 0$ and $g \neq 0$ the model is not analytically solvable, and the discrete symmetry group is reduced to
\begin{equation}
    \mathsf{G}_2 = \langle \sigma, \tau \rangle
\end{equation}
where
\begin{equation}
    \sigma \equiv \sigma_1 \circ \sigma_2 : (M_1, M_2) \mapsto (-M_1, -M_2) \, ,
    \quad
    \tau(M_1, M_2) = (M_2, M_1) \, .
\end{equation}
Consequently, loops of odd length vanish, e.g.\ $\langle \phi(M_1) \rangle = \langle \phi(M_1^3) \rangle = \langle \phi(M_1^2 M_2) \rangle = 0$. In contrast to the case $k = 0$, however, even-length loops with odd numbers of $M_1$ and of $M_2$ survive, e.g.\ $\langle \phi(M_1 M_2) \rangle \neq 0$. 

For $g = 0$ the model is again solvable, since the rotation
\begin{equation}
    M_{\pm} = \frac{1}{\sqrt{2}}(M_1 \pm M_2)
\end{equation}
diagonalizes the Hamiltonian,
\begin{equation}
    H = \tr(\frac{1}{2} P_{+}^2 + \frac{1}{2} P_{-}^2 + \frac{k_{+}}{2} M_{+}^2 + \frac{k_{-}}{2} M_{-}^2) \, ,
\end{equation}
where $k_{\pm} = 1 \pm k$. The ground state energy is
\begin{equation}
    E_{0} = \frac{1}{2} \bigl(\sqrt{k_{+}} + \sqrt{k_{-}}\bigr) \, ,
\end{equation}
and the lowest loops are
\begin{align}
    & \langle\phi(M_1^2)\rangle= \langle\phi(M_2^2)\rangle = \frac{1}{4} \bigl( \sqrt{1/k_{+}} + \sqrt{1/k_{-}} \bigr) \, , \\
    & \langle\phi(M_1 M_2)\rangle = \frac{1}{4} \bigl( \sqrt{1/k_{+}} - \sqrt{1/k_{-}} \bigr) \, .
\end{align}
Hence $k = \pm 1$ are critical values beyond which the quadratic Hamiltonian is unbounded from below, whereas for $g > 0$ the quartic terms stabilize the system for any $k$. 

\begin{figure}[!t]
\begin{center}
    \includegraphics[width=0.45\textwidth]{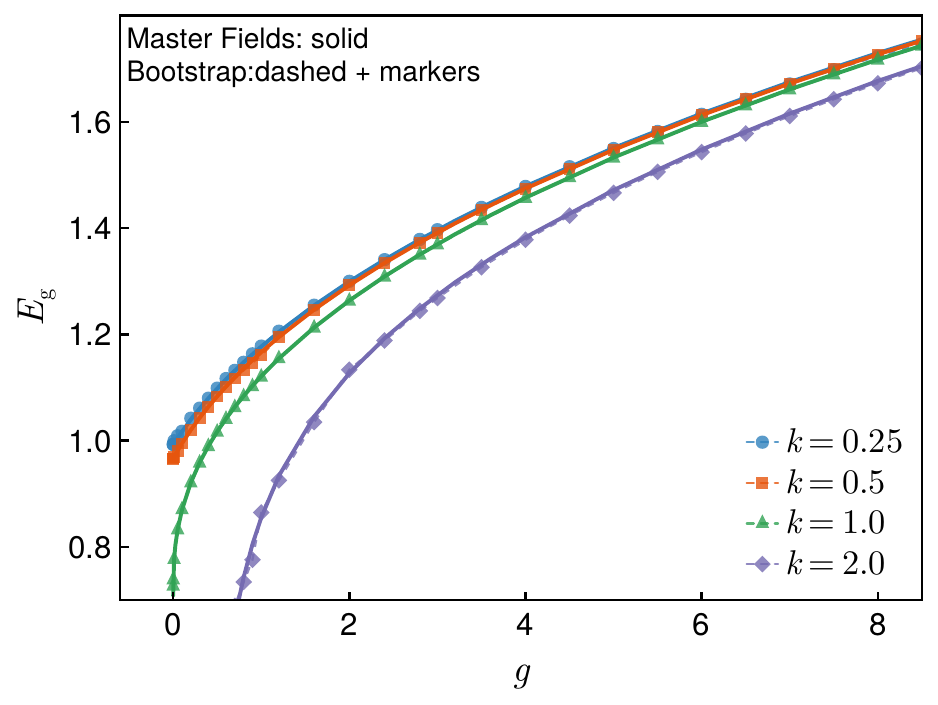}
    \caption{Ground state energy of the coupled two-matrix model versus $g$ for several $k \neq 0$, at truncation $\ell_{\max} = 8$. Markers and dashed lines are collective bootstrap results; solid lines are master field results with loop truncation $\ell_{\max}=6$ and master field size $N=16$.}
    \label{fig:2MQM_Eg_vs_g}
\end{center}
\end{figure}

\begin{figure}[!t]
\begin{center}
    \begin{subfigure}[t]{0.45\textwidth}
        \includegraphics[width=1\textwidth]{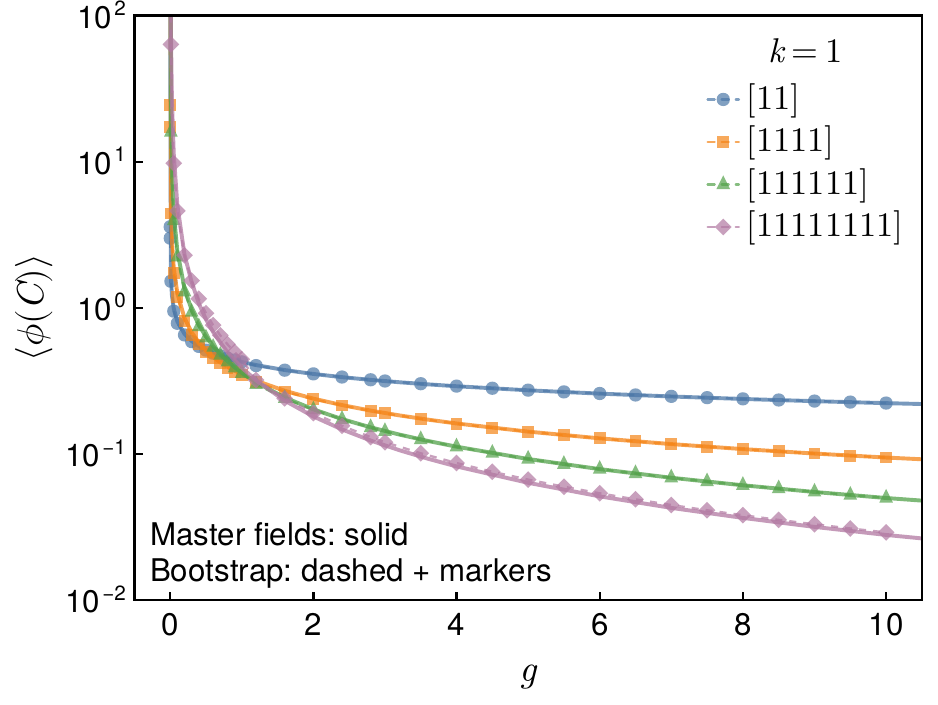}
    \end{subfigure}
    \begin{subfigure}[t]{0.45\textwidth}
        \includegraphics[width=1\textwidth]{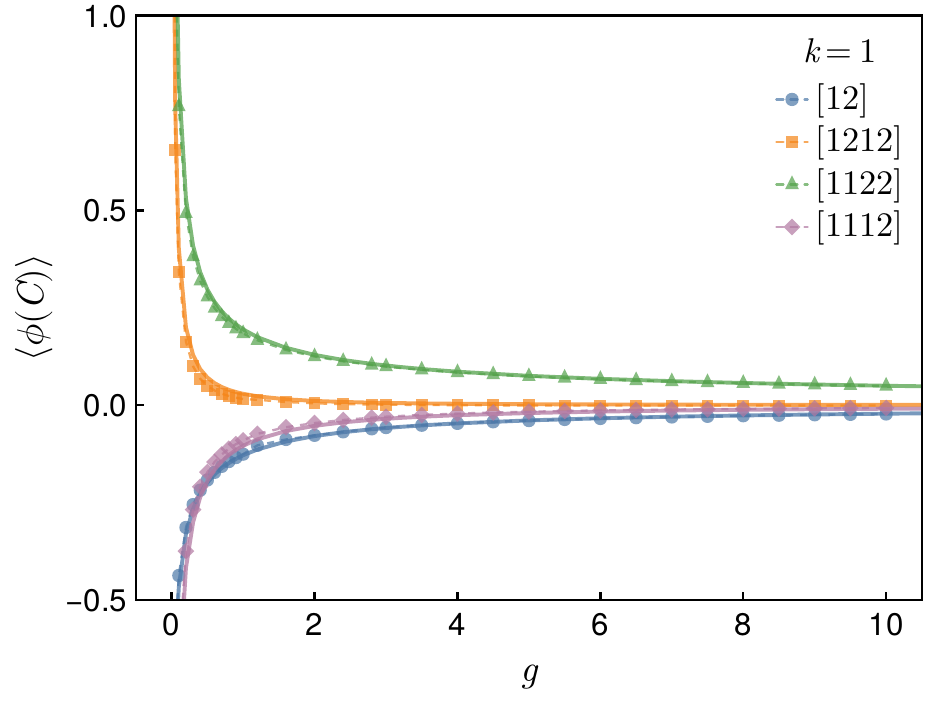}
    \end{subfigure}
    \caption{Loops of the types $\phi(C_1 C_2)$ (left) and $\phi(C_1 C_2 C_1' C_2')$ (right) for the coupled two-matrix model at $k=1$ and $\ell_{\max} = 8$. Markers and dashed lines are bootstrap results; solid lines are master field results with $\ell_{\max} = 6$ and $N=16$.} 
    \label{fig:2MQM_loop_k1_set1_and_2}
\end{center}
\end{figure}

At $\ell_{\max} = 8$ there are $N_{\rm{Loops}} = 2615$ loops, of which 560 are independent and nonzero after using $\mathsf{G}_2$. Since there are 14 such loops of length $\leq 6$, we introduce $14 \times 15 / 2 = 105$ lifted variables, for a total of 665 variables. The ground state energy for several values of $k$ is compared with master field optimization in \cref{fig:2MQM_Eg_vs_g}, and the loop expectation values for $k = 1$ are shown in \cref{fig:2MQM_loop_k1_set1_and_2}.

\section{Two-matrix Yang--Mills model}
\label{sec:two-MQM_Yang-Mills}

\subsection{Symmetries}

We next consider two Hermitian matrices with a Yang--Mills type interaction, studied with phase space bootstrap in \cite{Han:2020bkb} and with master field optimization in \cite{Mathaba:2023non},
\begin{equation}
    H = \tr(\frac{1}{2} P_1^2 + \frac{1}{2} P_2^2 + \frac{m^2}{2}M_1^2 + \frac{m^2}{2} M_2^2 - \frac{\lambda}{N}[M_1, M_2]^2) \, ,
\end{equation}
where $\lambda \equiv g_{\rm{YM}}^2$ is the 't Hooft coupling. 
Because of the $\rm{SO}(2) \cong \rm{U}(1)$ symmetry $M_i \rightarrow R_{ij} M_j$, $R \in \rm{SO}(2)$, it is convenient to work with
\begin{equation}
    Z \equiv M_1 + \ii M_2 \, , 
    \quad
    \bar{Z} \equiv Z^\dagger = M_1 - \ii M_2 \, ,
\end{equation}
so that only $\rm{U}(1)$ neutral loops with equal numbers of $Z$ and $\bar{Z}$ can have nonzero expectation values. Denoting by $P$ and $\bar{P}$ the momenta conjugate to $Z$ and $\bar{Z}$, the Hamiltonian reads
\begin{equation}
    H = \tr(2 P \bar{P} + \frac{m^2}{2} Z \bar{Z} + \frac{\lambda}{4 N} [Z, \bar{Z}]^2) \, .
\end{equation}
The $\rm{U}(1)$ charge is
\begin{equation}
    \mathcal{Q} = \tr(Z \partial_{Z} - \bar{Z} \partial_{\bar{Z}}) \, ,
    \quad
    [H, \mathcal{Q}] = 0 \, .
\end{equation}
In addition, there are two discrete symmetries:
\begin{enumerate}
    \item The reflection $M_2 \rightarrow - M_2$, which exchanges $Z \leftrightarrow \bar{Z}$ and implies $\langle \phi(C) \rangle = \langle \phi(\bar{C}) \rangle$, where $\bar{C}(Z, \bar{Z}) \equiv C(\bar{Z}, Z)$.
    \item The transposition $M_i \rightarrow M_i^{\rm{T}}$, which implies $\langle \phi(C) \rangle = \langle \phi(\operatorname{rev}(C)) \rangle$, where $\operatorname{rev}(C)$ denotes the reversed word.
\end{enumerate}
Consequently, the expectation values of all neutral loops are real:
\begin{equation}
    \overline{\langle \phi(C) \rangle} 
    = \langle \phi(C^\dagger) \rangle 
    = \langle \phi(\operatorname{rev}(\bar{C})) \rangle
    = \langle \phi(\bar{C}) \rangle
    = \langle \phi(C) \rangle \, . 
\end{equation}
In this basis the loop joining and loop splitting are
\begin{align}
    \Omega(C_1, C_2) & \equiv 2 N^2 \tr(
        \pdv{\phi(C_1^\dagger)}{Z} \pdv{\phi(C_2)}{\bar{Z}}
        +
        \pdv{\phi(C_1^\dagger)}{\bar{Z}} \pdv{\phi(C_2)}{Z}
    ) \, , \\
    \omega(C) & \equiv 4 \tr(
        \pdv[2]{\phi(C)}{Z}{\bar{Z}}
    ) \, .
\end{align}
Let $V_{q}$ denote the subspace spanned by independent loops of charge $q$. The loop joining maps $V_{q} \otimes V_{q'}$ to $V_{q'-q}$ while the loop splitting maps $V_{q}$ to $\bigoplus_{q_1} V_{q_1} \otimes V_{q - q_1}$. By charge conservation, the collective potential involves only the neutral sector,
\begin{equation}
    V_{\col} = \frac{1}{8} \omega_{0}^{\rm{T}} \Omega_{00}^{-1} \omega_0^{ } + \frac{m^2}{2} \phi(Z \bar{Z}) + \frac{\lambda}{4} \phi([Z, \bar{Z}]^2) \, ,
\end{equation}
where the subscripts refer to charge zero.
Using the virial theorem, the energy is
\begin{equation}
    \frac{1}{N^2} E  = m^2 \langle \phi(Z \bar{Z}) \rangle + \frac{3 \lambda}{4} \langle \phi([Z, \bar{Z}]^2) \rangle \, .
\end{equation}
Loops in the $\{M_1, M_2\}$ and $\{Z, \bar{Z} \}$ bases are related by linear transformations, which for the lowest lengths read 
%
\begin{equation}
    \begin{bmatrix}
    \tr Z \\  \tr Z^\dagger
    \end{bmatrix}
    =
    \begin{bmatrix}
        1 & \ii \\
        1 & - \ii
    \end{bmatrix}
    \begin{bmatrix}
        \tr M_1 \\ \tr M_2
    \end{bmatrix} \, .
\end{equation}
%
\begin{equation}
    \begin{bmatrix}
        \tr Z^2 \\ \tr Z Z^\dagger \\ \tr Z^{\dagger 2}
    \end{bmatrix}
    =
    \begin{bmatrix}
        1 & 2 \ii & -1 \\
        1 & 0 & 1 \\
        1 & -2 \ii & -1 \\
    \end{bmatrix}
    \begin{bmatrix}
        \tr M_1^2 \\ \tr M_1 M_2 \\ \tr M_2^2
    \end{bmatrix} \, .
\end{equation}
%
\begin{equation}
    \begin{bmatrix}
        \tr Z^3 \\ \tr Z^2 Z^\dagger \\ \tr Z Z^{\dagger 2} \\ \tr Z^{\dagger 3}
    \end{bmatrix}
    =
    \begin{bmatrix}
        1 & 3 \ii & -3 & -\ii \\
        1 & \ii & 1 & \ii \\
        1 & -\ii & 1 & -\ii \\
        1 & -3 \ii & -3 & \ii \\
    \end{bmatrix}
    \begin{bmatrix}
        \tr M_1^3 \\ \tr M_1^2 M_2 \\ \tr M_1 M_2^{2} \\ \tr M_2^{3}
    \end{bmatrix} \, .
\end{equation}
%
\begin{equation}
    \begin{bmatrix}
        \tr Z^4 \\ \tr Z^3 Z^\dagger \\ \tr Z^2 Z^{\dagger 2} \\ \tr Z Z^\dagger Z Z^\dagger \\ \tr Z Z^{\dagger 3} \\ \tr Z^{\dagger 4}
    \end{bmatrix}
    =
    \begin{bmatrix}
        1 & 4 \ii & -4 & -2 & -4 \ii & 1 \\
        1 & 2 \ii & 0 & 0 & 2 \ii & -1 \\
        1 & 0 & 0 & 2 & 0 & 1 \\
        1 & 0 & 4 & -2 & 0 & 1 \\
        1 & -2 \ii & 0 & 0 & -2 \ii & -1 \\
        1 & -4 \ii & -4 & -2 & 4 \ii & 1 \\
    \end{bmatrix}
    \begin{bmatrix}
        \tr M_1^4 \\ \tr M_1^3 M_2 \\ \tr M_1^2 M_2^{2} \\ \tr M_1 M_2 M_1 M_2 \\ \tr M_1 M_2^{3} \\ \tr M_2^{4}
    \end{bmatrix} \, .
\end{equation}
%
\begin{equation}
    \begin{bmatrix}
        \tr Z^5 \\ 
        \tr Z^4 Z^\dagger \\ 
        \tr Z^3 Z^{\dagger 2} \\ 
        \tr Z^2 Z^\dagger Z Z^\dagger \\ 
        \tr Z^2 Z^{\dagger 3} \\
        \tr Z Z^\dagger Z Z^{\dagger 2} \\ 
        \tr Z Z^{\dagger 4} \\ 
        \tr Z^{\dagger 5}
    \end{bmatrix}
    =
    \begin{bmatrix}
        1 & 5 \ii & -5 & -5 & -5 \ii & -5 \ii & 5 & \ii \\
        1 & 3 \ii & -1 & -1 & \ii & \ii & -3 & -\ii \\
        1 & \ii & -1 & 3 & -\ii & 3 \ii & 1 & \ii \\
        1 & \ii & 3 & -1 & 3 \ii & -\ii & 1 & \ii \\
        1 & -\ii & -1 & 3 & \ii & -3 \ii & 1 & -\ii \\
        1 & -\ii & 3 & -1 & -3 \ii & \ii & 1 & -\ii \\
        1 & -3 \ii & -1 & -1 & -\ii & -\ii & -3 & \ii \\
        1 & -5 \ii & -5 & -5 & 5 \ii & 5 \ii & 5 & -\ii \\
    \end{bmatrix}
    \begin{bmatrix}
        \tr M_1^5 \\
        \tr M_1^4 M_2 \\
        \tr M_1^3 M_2^2 \\
        \tr M_1^2 M_2 M_1 M_2 \\
        \tr M_1^2 M_2^3 \\
        \tr M_1 M_2 M_1 M_2^2 \\
        \tr M_1 M_2^4 \\
        \tr M_2^5
    \end{bmatrix} \, .
\end{equation}

\subsection{Master field optimization}

The master field optimization of \cite{Mathaba:2023non} was performed in the $\{M_1, M_2\}$ basis. Since $M_1$ and $M_2$ cannot be simultaneously diagonalized, that work used $N$ real variables for the eigenvalues of $M_1$ and $N^2$ real variables for $M_2$, i.e.\ $N + N^2$ parameters in total. To provide an independent benchmark for the collective bootstrap, we instead work in the $\{Z, \bar{Z}\}$ basis with a parametrization that respects the symmetries, which reduces the number of master field variables. We take the space on which the master field acts to be graded by the $\rm{U}(1)$ charge, $\mathcal{H} = \oplus_{q} \mathcal{H}_{q}$. Then $Z$ and $\bar{Z}$ only connect neighboring sectors $\mathcal{H}_{q}$ and $\mathcal{H}_{q+1}$, shifting the grading in opposite directions, so that we can think of them as creation and annihilation operators. Motivated by the transposition symmetry, we take $Z$ to be real, so that $\bar{Z} = Z^{\dagger} = Z^{\rm{T}}$. We emphasize that this is a choice of ansatz rather than a consequence of the symmetry. Moreover, the $\mathbb{Z}_2$ symmetry $Z \leftrightarrow \bar{Z}$ is implemented by mapping $\mathcal{H}_{q}$ to $\mathcal{H}_{-q}$. Taking these symmetries into account, we parametrize $Z$ as a block superdiagonal matrix
\begin{equation}
    Z = \bordermatrix{
  q \backslash q' & \cdots & -2 & -1 & 0 & 1 & 2 & \cdots \cr
  \,\, \vdots          & \ddots & \vdots  & \vdots & \vdots & \vdots & \vdots & \iddots \cr
  -2         & \cdots & 0  & Z_{1,2}^{\rm{T}} & 0 & 0 & 0 & \cdots \cr
  -1              & \cdots & 0  & 0 & Z_{0,1}^{\rm{T}} & 0 & 0 & \cdots \cr
  \,\, 0              & \cdots & 0  & 0 & 0 & Z_{0,1}      & 0 & \cdots \cr
  \,\, 1               & \cdots & 0  & 0 & 0 & 0 & Z_{1,2}  & \cdots \cr
  \,\, 2               & \cdots & 0  & 0 & 0 & 0      & 0 & \cdots \cr
  \,\, \vdots          & \iddots & \vdots  & \vdots & \vdots & \vdots & \vdots & \ddots \cr
} \, ,
\end{equation}
and $\bar{Z} = Z^{\rm{T}}$. Each $Z_{q,q+1}$ connects $\mathcal{H}_{q}$ and $\mathcal{H}_{q+1}$ and is a real block matrix of size $\dim(\mathcal{H}_{q}) \times \dim(\mathcal{H}_{q+1})$. With this parametrization all loops with nonzero $\rm{U}(1)$ charge vanish identically, and the neutral loops are easily computed. For example, denoting by $D$ the total size of $Z$,
\begin{align}
    \phi(Z \bar{Z}) & = \frac{1}{D^2} \sum_{q} \tr(Z_{q, q+1} \bar{Z}_{q+1, q}) \, , \\
    \phi(Z Z \bar{Z} \bar{Z}) & = \frac{1}{D^3} \sum_{q} \tr(Z_{q, q+1} Z_{q+1, q+2} \bar{Z}_{q+2, q+1} \bar{Z}_{q+1, q}) \, , \\
     \phi(Z \bar{Z} Z \bar{Z}) & = \frac{1}{D^3} \sum_{q} \tr(Z_{q, q+1} \bar{Z}_{q+1, q} Z_{q, q+1} \bar{Z}_{q+1, q}) \, .
\end{align}
Master field results obtained with this parametrization at loop truncation $\ell_{\max} = 6$ are compared with the collective bootstrap below. 

\subsection{Collective bootstrap}

The SDPs are solved with \texttt{MOSEK}, and the tolerance of the rank minimization is $\lambda_2 / \lambda_1 < 10^{-7}$. We truncate the loop space at $\ell_{\max} = 8$, so that $N_{\Omega} = 93$ and $N_{\rm{Loops}}=2615$. Using the symmetries, $146$ independent loops can be nonzero. Since only $6$ of them have length $\leq \ell_{\max} - 2 = 6$, we introduce $6 \times 7 / 2 = 21$ lifted variables $Y$.

\subsubsection{Massless case}

\begin{figure}[!thb]
\begin{center}
    \includegraphics[width=0.5\textwidth]{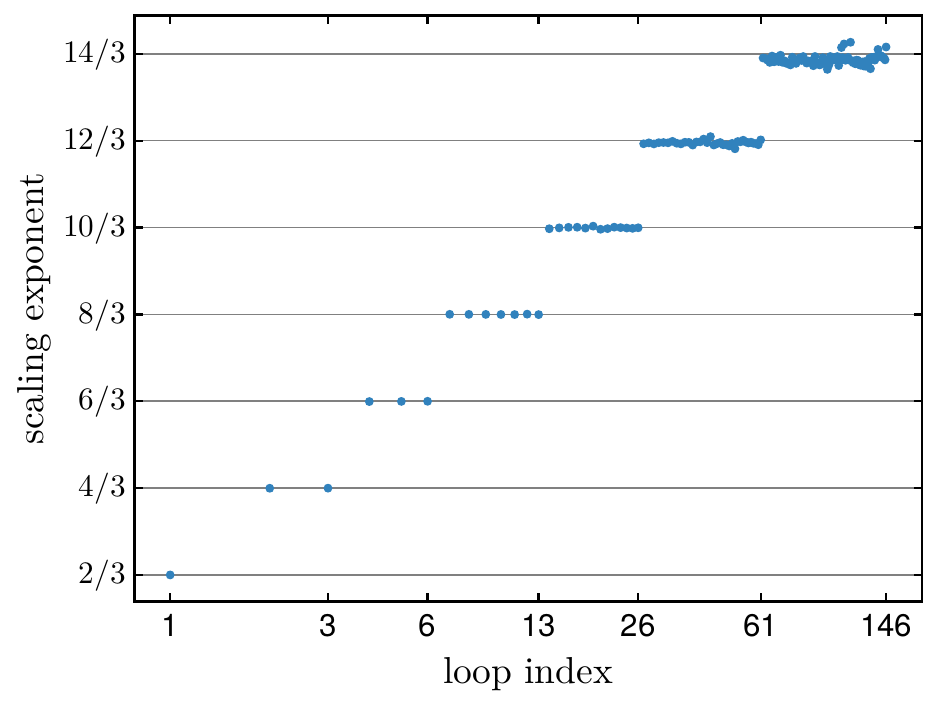}
    \caption{Fitted scaling exponents $p_C$ of the loops for the massless two-matrix Yang--Mills model, compared with the exact value $\len(C)/3$.}
    \label{fig:2MQM_YM_m0_scaling}
\end{center}
\end{figure}

\begin{figure}[!thb]
\begin{center}
    \includegraphics[width=0.5\textwidth]{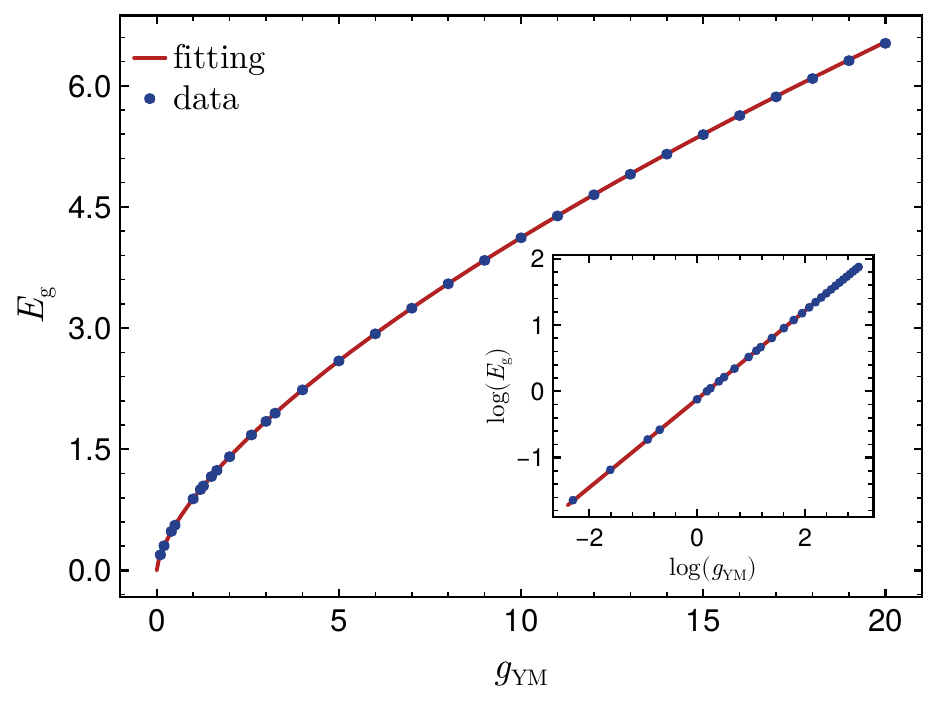}
    \caption{Ground state energy of the massless two-matrix Yang--Mills model versus $g_{\rm YM}$. The inset shows the fit to $\Lambda_E \, g_{\rm YM}^{2/3}$.}
    \label{fig:2MQM_YM_m0_E_vs_g}
\end{center}
\end{figure}

\begin{figure}[!thb]
\begin{center}
    \begin{subfigure}[t]{0.45\textwidth}
        \includegraphics[width=1\textwidth]{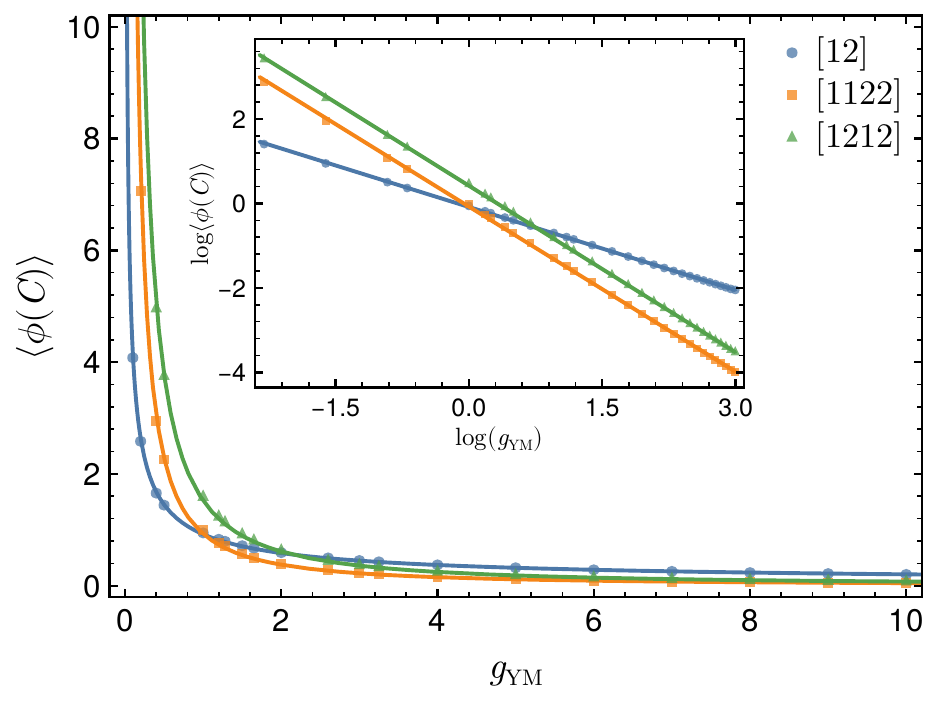}
    \end{subfigure}
    \begin{subfigure}[t]{0.45\textwidth}
        \includegraphics[width=1\textwidth]{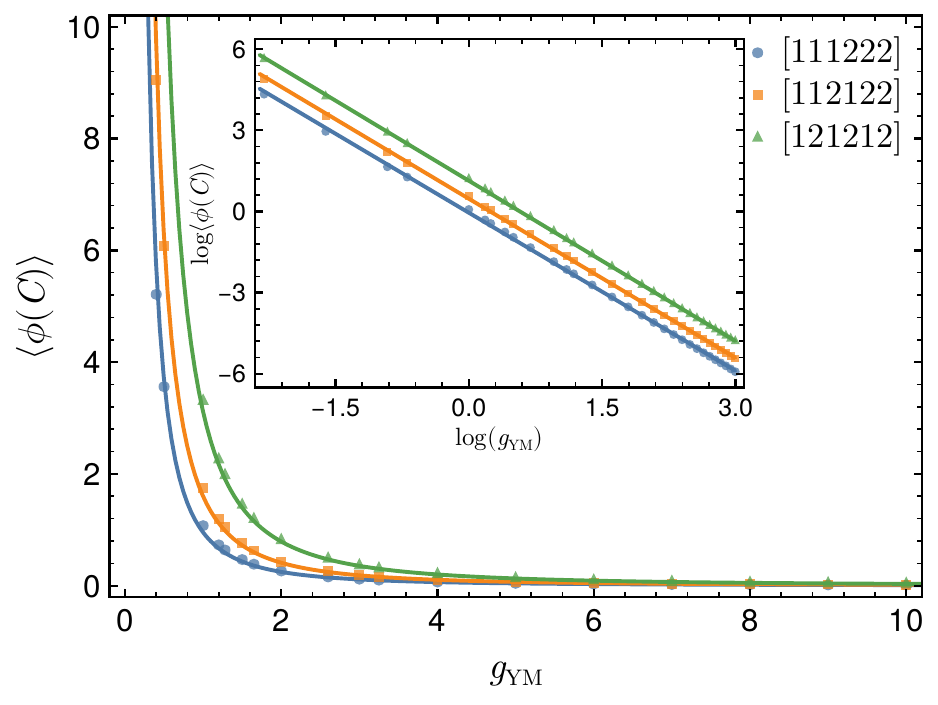}
    \end{subfigure}
    \begin{subfigure}[t]{0.45\textwidth}
        \includegraphics[width=1\textwidth]{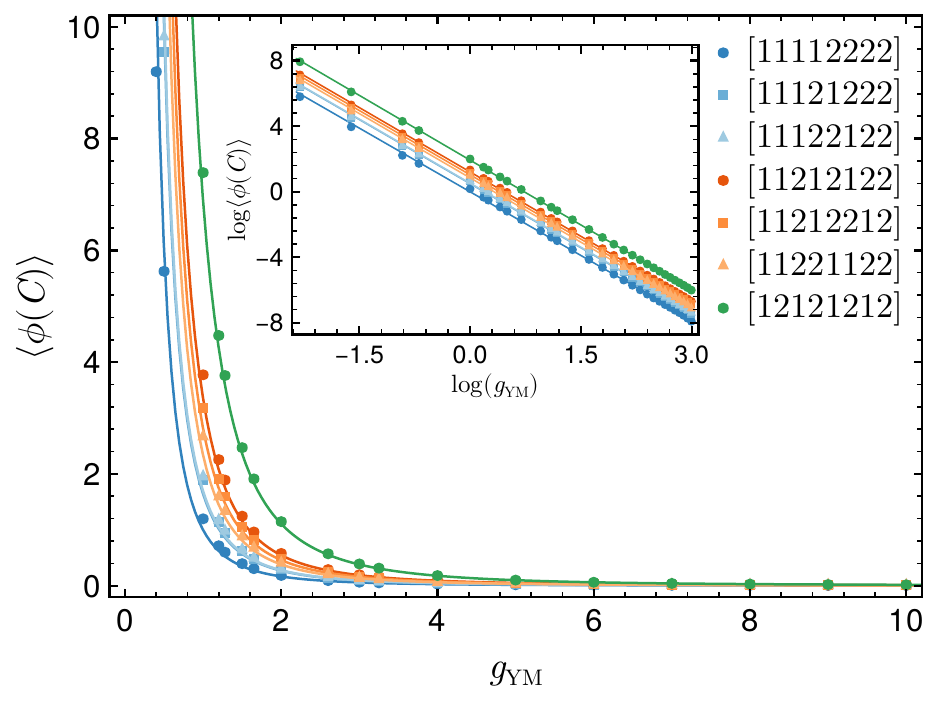}
    \end{subfigure}
    \begin{subfigure}[t]{0.45\textwidth}
        \includegraphics[width=1\textwidth]{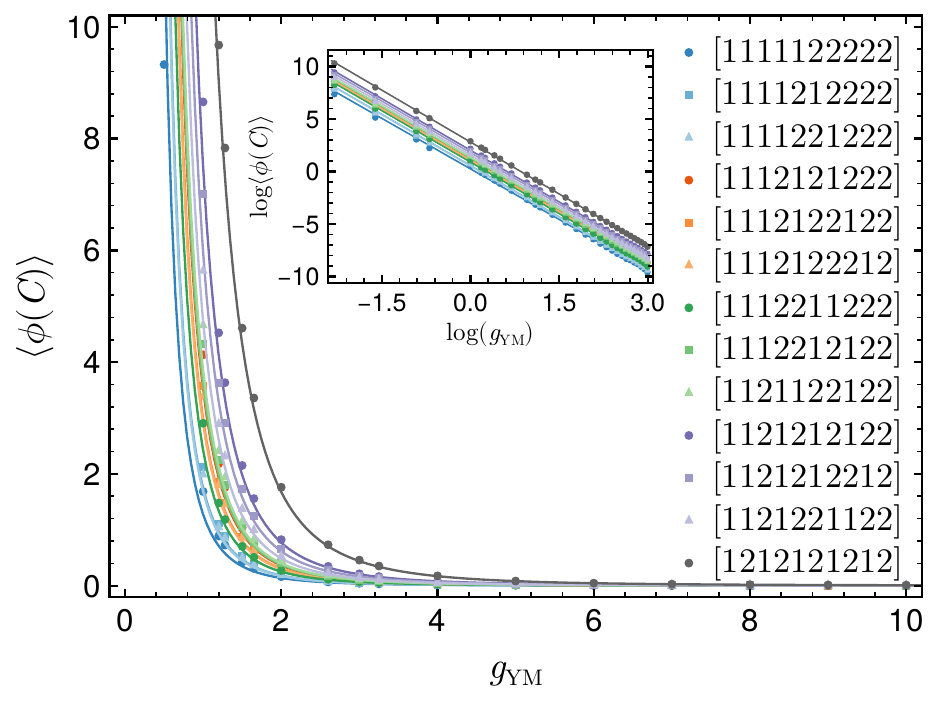}
    \end{subfigure}
    \caption{Loop expectation values of the massless two-matrix Yang--Mills model. Markers are bootstrap results and solid lines are fits to the scaling form \eqref{eq:2MQM_YM_scaling}.}
    \label{fig:2MQM_YM_m0_loop_vs_g}
\end{center}
\end{figure}

For $m=0$ we restrict to traceless Hermitian matrices, since the trace parts would otherwise be free particles without a normalizable ground state. Although the classical potential has flat directions along commuting matrices, the quantum spectrum admits a well-defined ground state. The only parameter is $g_{\rm{YM}}$, and rescaling $M_a \to g_{\rm YM}^{-1/3} M_a$, $P_a \to g_{\rm YM}^{1/3} P_a$ implies the exact scaling \cite{Han:2020bkb,Mathaba:2023non}
\begin{equation} \label{eq:2MQM_YM_scaling}
    E_{\rm{g}} = \Lambda_{E} \, g_{\rm{YM}}^{2/3} \, ,
    \quad 
    \langle \phi(C) \rangle =  \Lambda_{C} \, g_{\rm{YM}}^{- \len(C) / 3} \, .
\end{equation}
As a check of the numerics, we fit the loop expectation values to
\begin{equation} \label{eq:2MQM_m0_fit_formula}
    \ln \langle \phi(C) \rangle = - p_C \ln g_{\rm{YM}} + \ln \Lambda_{C} \, ,
\end{equation}
for which we expect $p_C = \len(C) / 3$. The fitted exponents are shown in \cref{fig:2MQM_YM_m0_scaling}. Moreover, we compare the loop expectation values at $g = 1$ with the fitted $\Lambda_{C}$ in \cref{tab:2MQM_YM_m0_loop_values}. Since the scaling \eqref{eq:2MQM_YM_scaling} follows exactly from rescaling the matrices, the two columns would coincide for exact data, and their difference measures the numerical uncertainty of the loops. For loops of length 2 and 4 the difference is about $1\%$ or less, it grows to $2$--$3\%$ at length 8, and reaches about $6\%$ for some loops of length 10 (e.g.\ $[1112211222]$). This is consistent with the expectation of \cref{subsubsec:status_of_results} that loops longer than $\ell_{\max}$ are only weakly constrained. The ground state energy and the loop expectation values are shown in \cref{fig:2MQM_YM_m0_E_vs_g,fig:2MQM_YM_m0_loop_vs_g}, where the insets display the fits \eqref{eq:2MQM_m0_fit_formula} and the analogous fit for the energy. The fit gives
\begin{equation}
    E_{\rm{g}} \approx 0.887 g_{\rm{YM}}^{2/3} \, ,
\end{equation}
close to the master field value $E^{\rm{MF}} \approx 0.8890(2) g_{\rm{YM}}^{2/3}$ in \cite{Mathaba:2023non}.
The commutator $\langle\tr([M_1, M_2]^2)\rangle$ does not vanish, and the dimensionless ratio \cite{Han:2020bkb,Mathaba:2023non}
\begin{equation}
    \theta \equiv - \frac{N}{2} \frac{\langle \tr([M_1, M_2]^2) \rangle}{\langle \tr(M_1^2) \rangle \langle \tr(M_2^2) \rangle}
    =
    \frac{\langle \phi(Z\bar{Z}Z\bar{Z}) - \phi(ZZ\bar{Z}\bar{Z}) \rangle}{\langle \phi(Z\bar{Z}) \rangle ^2} \approx 0.671 \, ,
\end{equation}
is independent of the coupling. Since at $m = 0$ the virial theorem gives $E = 3 \langle V \rangle$, the energy and the angle are related by $E_{\rm g} = \frac{3 \lambda}{2} \theta \langle \phi(Z \bar{Z}) \rangle^2$. With the values above, the right hand side gives $0.886$, consistent with the fitted $\Lambda_E = 0.887$.

\begin{table}[htbp]
\centering
\caption{Loop expectation values of the massless two-matrix Yang--Mills model at $g_{\rm YM}=1$, and the prefactors $\Lambda_C$ obtained from fits over the whole range of couplings. In the words, $1 \equiv Z$ and $2 \equiv \bar{Z}$.}
\label{tab:2MQM_YM_m0_loop_values}
  \begin{tabular}{ccc  @{\hspace{5em}}  ccc}
    \toprule
    Word & Value ($g=1$) &  $\Lambda_{C}$ Fit & Word & Value ($g=1$) &  $\Lambda_{C}$ Fit \\
    \midrule
    $[12]$ & 0.9384 & 0.9444 & $[1111122222]$ & 1.6338 & 1.6850 \\
    $[1122]$ & 0.9828 & 0.9952 & $[1111212222]$ & 2.0422 & 2.1235 \\
    $[1212]$ & 1.5742 & 1.5860 & $[1111221222]$ & 1.9145 & 1.9923 \\
    $[111222]$ & 1.0549 & 1.0742 & $[1112121222]$ & 4.0091 & 4.1309 \\
    $[112122]$ & 1.7268 & 1.7512 & $[1112122122]$ & 3.4094 & 3.5688 \\
    $[121212]$ & 3.2549 & 3.3034 & $[1112122212]$ & 3.3006 & 3.3891 \\
    $[11112222]$ & 1.1655 & 1.1966 & $[1112211222]$ & 2.7352 & 2.9057 \\
    $[11121222]$ & 1.8476 & 1.8930 & $[1112212122]$ & 4.1418 & 4.3306 \\
    $[11122122]$ & 1.9096 & 1.9525 & $[1121122122]$ & 4.4348 & 4.6625 \\
    $[11212122]$ & 3.6787 & 3.7737 & $[1121212122]$ & 8.3278 & 8.6564 \\
    $[11212212]$ & 3.1053 & 3.1738 & $[1121212212]$ & 6.6947 & 7.0105 \\
    $[11221122]$ & 2.5968 & 2.6703 & $[1121221122]$ & 5.3448 & 5.6359 \\
    $[12121212]$ & 7.2945 & 7.3907 & $[1212121212]$ & 17.8056 & 17.7000 \\
    \bottomrule
  \end{tabular}
\end{table}

\subsubsection{Massive case}

\begin{figure}[!thb]
\begin{center}
    \includegraphics[width=0.5\textwidth]{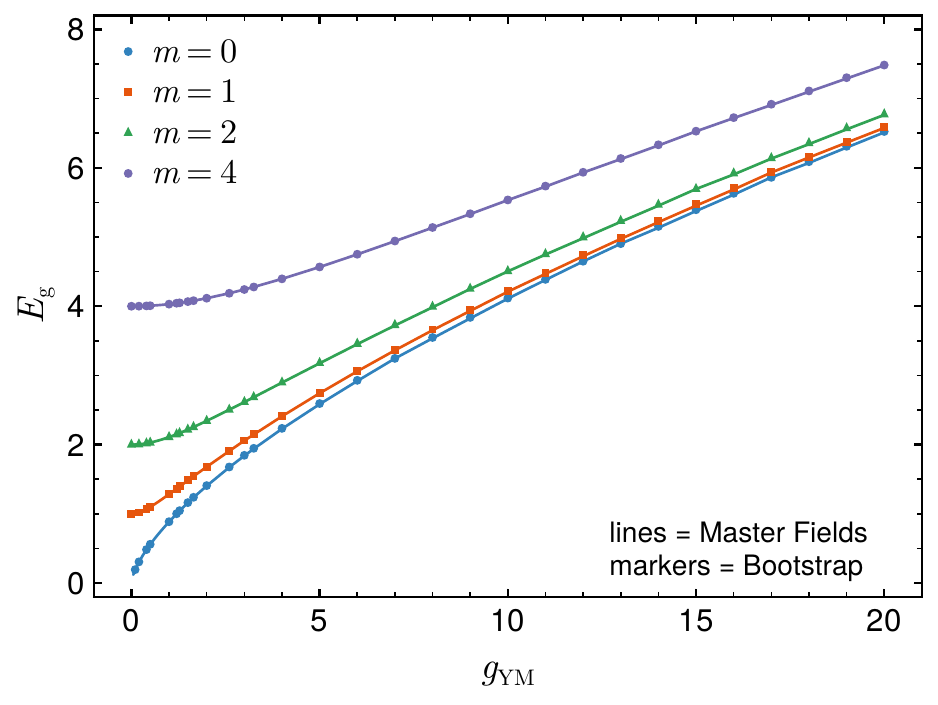}
    \caption{Ground state energy of the massive two-matrix Yang--Mills model versus $g_{\rm YM}$ for several masses (scatters), compared with master field optimization (lines).}
    \label{fig:Eg_2MQM_YM_m_neq0}
\end{center}
\end{figure}

\begin{figure}[!thb]
\begin{center}
    \begin{subfigure}[t]{0.45\textwidth}
        \includegraphics[width=1\textwidth]{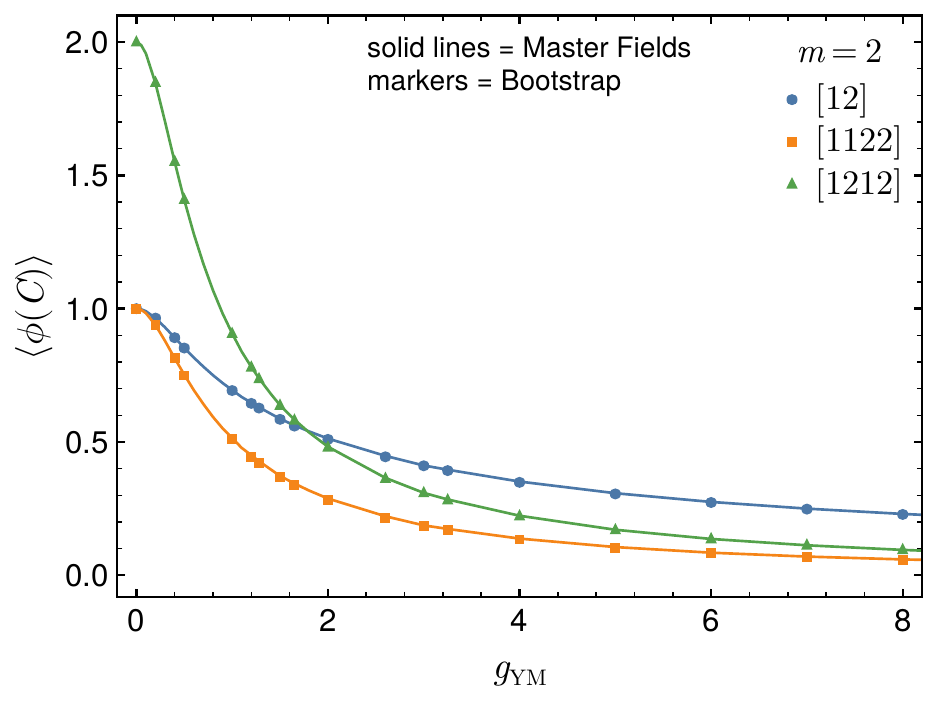}
    \end{subfigure}
    \begin{subfigure}[t]{0.45\textwidth}
        \includegraphics[width=1\textwidth]{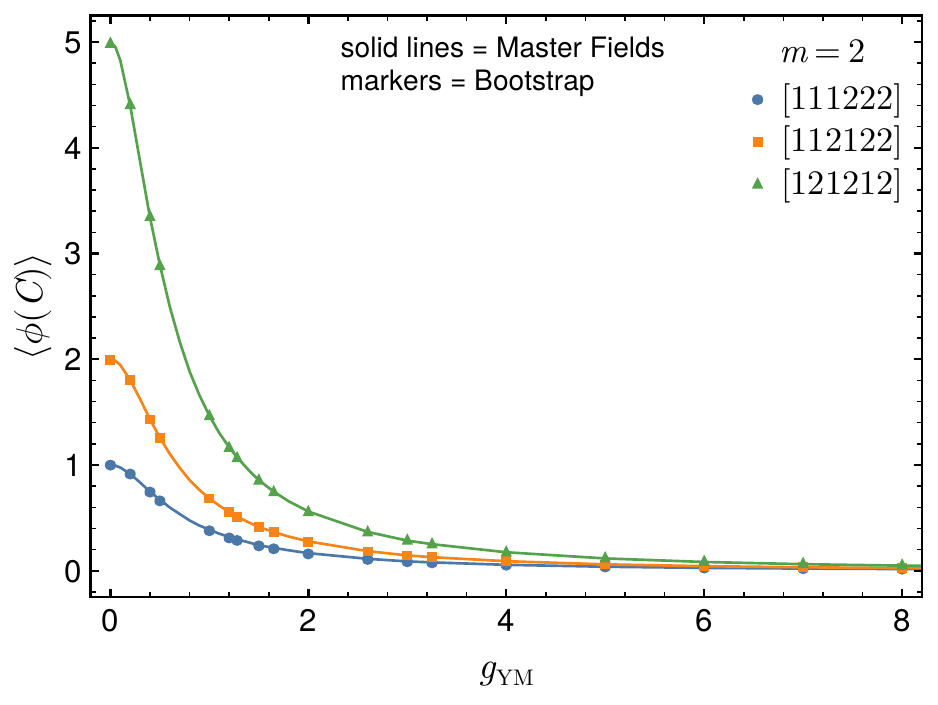}
    \end{subfigure}
    \begin{subfigure}[t]{0.45\textwidth}
        \includegraphics[width=1\textwidth]{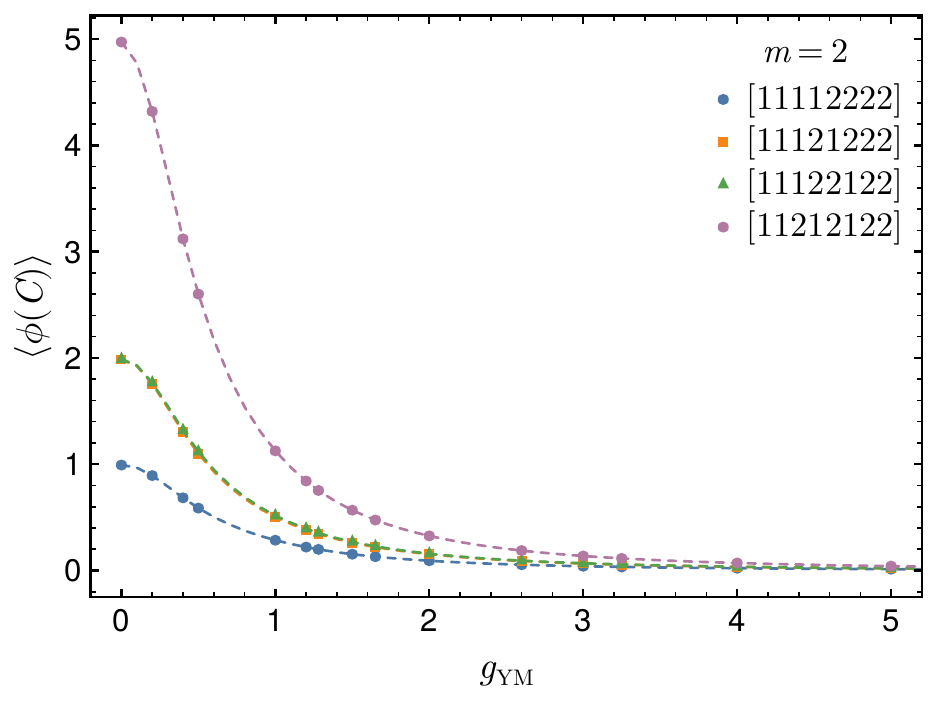}
    \end{subfigure}
    \begin{subfigure}[t]{0.45\textwidth}
        \includegraphics[width=1\textwidth]{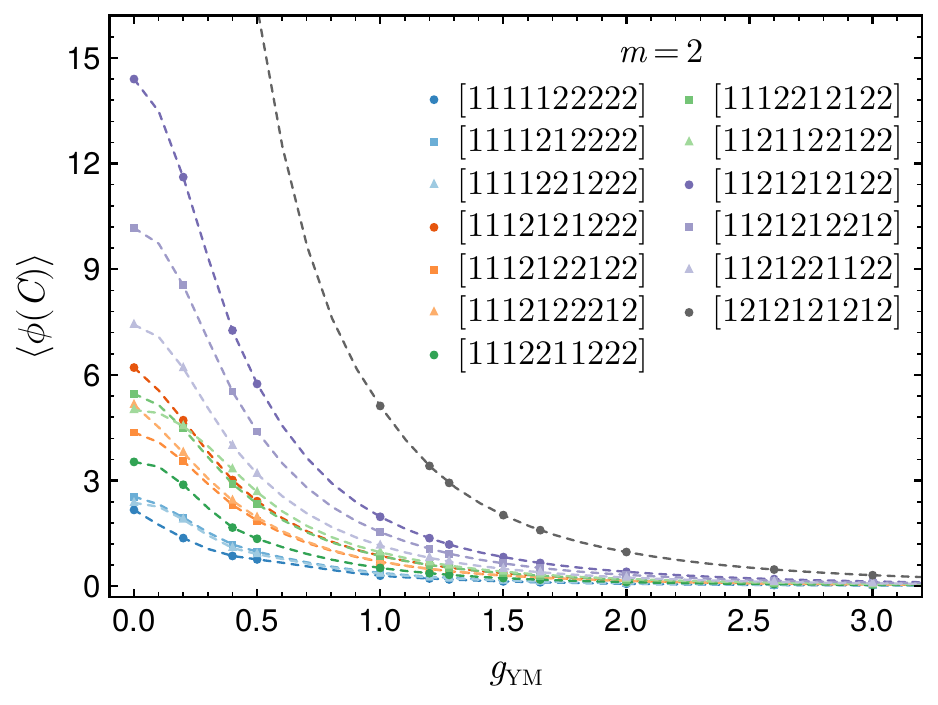}
    \end{subfigure}
    \caption{Loop expectation values of the two-matrix Yang--Mills model at $m = 2$. In the top panels, collective bootstrap results (markers) and master field results (solid lines) are compared. In the bottom panels, markers are collective bootstrap results, connected by spline interpolations (dashed lines) to guide the eye.}
    \label{fig:loop_2MQM_YM_m2}
\end{center}
\end{figure}

For $m \neq 0$ we compare the bootstrap with master field optimization. The ground state energy is shown in \cref{fig:Eg_2MQM_YM_m_neq0}, and the loops at $m = 2$ in \cref{fig:loop_2MQM_YM_m2}. In the strong coupling regime $g_{\rm{YM}}^{2/3} \gg m$, a fit gives
\begin{equation}
    E_{\rm{g}} = 0.887 g_{\rm{YM}}^{2/3} + 0.455 \frac{m^2}{g_{\rm{YM}}^{2/3}} + \dots
\end{equation}
The coefficient of $m^2$ can be determined independently. By the Hellmann--Feynman theorem, $\partial E_{\rm g} / \partial m^2 = \langle \phi(Z \bar{Z}) \rangle  / 2$. Evaluating the right hand side at $m = 0$ with the scaling \eqref{eq:2MQM_YM_scaling} gives the coefficient $\frac{1}{2} \Lambda_{[12]}$. Using the value $\langle \phi(Z \bar{Z}) \rangle = 0.9384$ at $g_{\rm{YM}} = 1$ (or the fitted $\Lambda_{[12]} = 0.9444$) from \cref{tab:2MQM_YM_m0_loop_values}, this gives $0.469$ (or $0.472$). The fitted coefficient $0.455$ is about $3\%$ smaller, which we attribute to higher-order terms in $m^2 / g_{\rm{YM}}^{4/3}$ within the fitting window. 

\clearpage
\newpage
\section{Three-matrix Yang--Mills model}
\label{sec:three-MQM_Yang-Mills}

Finally, we consider three traceless Hermitian matrices with a massless Yang--Mills interaction,
\begin{equation}
    H = \operatorname{tr}\Biggl(\frac{1}{2} \sum_{a=1}^3 P_a^2 
    - \frac{g_{\rm{YM}}^2}{4N} \sum_{\substack{a, b = 1 \\ a \neq b}}^{3}[M_a, M_b]^2\Biggr) \, .
\end{equation}
This is the bosonic part of the dimensional reduction of four-dimensional Yang--Mills theory, and also serves as a step towards the nine-matrix bosonic sector of the BFSS model. As in the two-matrix case, the rescaling $M_a \rightarrow g_{\rm{YM}}^{-1/3} M_a$, $P_a \rightarrow g_{\rm{YM}}^{1/3} P_a$ gives an overall scaling $g_{\rm{YM}}^{2/3}$ of the Hamiltonian. The ground state energy thus scales as $g_{\rm{YM}}^{2/3}$ and loops of length $\ell$ as $g_{\rm YM}^{-\ell/3}$.
The Hamiltonian is invariant under $M_a \rightarrow R_{ab} M_b$ for all $R_{ab} \in \mathrm{O}(3)$, and under the transposition $M_a \rightarrow M_a^{\rm{T}}$. As a result, the relevant loop space is spanned by $\mathrm{O}(3)$-invariant words, which are built by pairwise contraction of flavor indices and taken modulo cyclicity and reversal. We use the following notation: letters $a, b, c, \dots$ denote flavor indices summed independently over $1, 2, 3$, and each letter appears exactly twice in a word. In particular, all loops of odd length vanish due to symmetries. For example, the first three loops are
\begin{gather}
    \phi(aa) \equiv \frac{1}{N^2} \sum_{a=1}^{3} \tr(M_a^2) \, ,
    \nonumber \\
    \phi(aabb) \equiv \frac{1}{N^3} \sum_{a,b=1}^{3} \tr(M_a^2 M_b^2) \, , 
    \quad
    \phi(abab) \equiv \frac{1}{N^3} \sum_{a,b=1}^{3} \tr(M_a M_b M_a M_b) \, .
\end{gather}
We assume an $\mathrm{O}(3)$-invariant, transpose-even extremal planar ground state. \Cref{tab:O3_loops} lists the 22 nontrivial invariant loops through length 8, and \cref{tab:O3_invariant} summarizes the number of loops at different truncations.
\begin{table}[htbp]
\centering
\caption{$\mathrm{O}(3)$-invariant, transpose-even loops through length 8, labeled by the loop index.}
\label{tab:O3_loops}
\begin{tabular}{cl @{\hspace{5em}} cl}
    \toprule
    loop index & loop word & loop index & loop word \\
    \midrule
    1 & $[a \, a]$ 
        & 12 & $[a \, a \, b \, c \, b \, d \, c \, d]$ \\
    2 & $[a \, a \, b \, b]$
        & 13 & $[a \, a \, b \, c \, b \, d \, d \, c]$\\
    3 & $[a \, b \, a \, b]$
        & 14 & $[a \, a \, b \, c \, d \, b \, c \, d]$\\
    4 & $[a \, a \, b \, b \, c \, c]$
        & 15 & $[a \, a \, b \, c \, d \, b \, d \, c]$\\
    5 & $[a \, a \, b \, c \, b \, c]$
        & 16 & $[a \, a \, b \, c \, d \, d \, b \, c]$\\
    6 & $[a \, a \, b \, c \, c \, b]$
        & 17 & $[a \, a \, b \, c \, d \, c \, d \, b]$\\
    7 & $[a \, b \, a \, c \, b \, c]$
        & 18 & $[a \, a \, b \, c \, d \, d \, c \, b]$\\
    8 & $[a \, b \, c \, a \, b \, c]$
        & 19 & $[a \, b \, a \, b \, c \, d \, c \, d]$\\
    9 & $[a \, a \, b \, b \, c \, c \, d \, d]$
        & 20 & $[a \, b \, a \, c \, b \, d \, c \, d]$\\
    10 & $[a \, a \, b \, b \, c \, d \, c \, d]$
        & 21 & $[a \, b \, a \, c \, d \, b \, c \, d]$\\
    11 & $[a \, a \, b \, b \, c \, d \, d \, c]$
        & 22 & $[a \, b \, a \, c \, d \, b \, d \, c]$\\
    \bottomrule
\end{tabular}
\end{table}
\begin{table}[htbp]
  \centering
  \caption{Number of independent $\mathrm{O}(3)$-invariant loops in the loop splitting ($N_{\omega}$), in the loop joining matrix ($N_{\Omega}$), and in total ($N_{\rm{Loops}}$) at different truncations. Here $\ell_{\max}$ is the truncation of the loop joining, and $L = 2 \ell_{\max} - 2$ is the maximal length of loops entering $\Omega$.}
  \label{tab:O3_invariant}
  \begin{tabular}{ccccc}
    \toprule
    $\ell_{\max}$ & $L$ & $N_{\omega}$ & $N_{\Omega}$ & $N_{\rm{Loops}}$ \\
    \midrule
    4 & 6 & 1 & 3 & 8 \\
    5 & 8 & 1 & 3 & 22 \\
    6 & 10 & 3 & 8 & 73 \\
    7 & 12 & 3 & 8 & 305 \\
    8 & 14 & 8 & 22 & 1544 \\
    \bottomrule
  \end{tabular}
\end{table}
In terms of these loops, labeled $\phi_i$ by the index in \cref{tab:O3_loops}, the loop joining matrix is
\begingroup\small
\begin{equation}
    \Omega = 
    \bordermatrix{
        \text{indices} & 1 & 2 & 3 & 4 & 5 & \cdots
        \cr
        1 & 4 \phi_1 & 8 \phi_2 & 8 \phi_3 & 12 \phi_4  & 12 \phi_5 & \cdots 
        \cr
        2 & 8 \phi_2 &  8 \phi_4 + 8 \phi_6 & 16 \phi_5 
        & 12 \phi_9 + 12 \phi_{11} 
        & 8 \phi_{10} + 8 \phi_{13} + 4 \phi_{16} + 4 \phi_{17} 
        & \cdots 
        \cr
        3 
        & 8 \phi_3 
        & 16 \phi_5 
        & 16 \phi_7
        & 24 \phi_{10}
        &  8 \phi_{12} + 8 \phi_{15} + 8 \phi_{19}
        & \cdots
        \cr
        4
        & 12 \phi_4 
        & 12 \phi_9 + 12 \phi_{11} 
        & 24 \phi_{10}
        & 18 \phi_{24} + 18 \phi_{32}
        & 12 \phi_{25} + 12 \phi_{28} + 6 \phi_{31} + 6 \phi_{33}
        & \cdots 
        \cr
        \vdots 
        & \vdots & \vdots & \vdots & \vdots & \vdots & \ddots
        \cr
    } \, ,
\end{equation}
\endgroup
where the length 10 loops are $\phi_{24} \equiv [a a b b c c d d e e]$, $\phi_{25} \equiv [a a b b c c d e d e]$, $\phi_{28} \equiv [a a b b c d c e e d]$, $\phi_{31} \equiv [a a b b c d e e c d]$, $\phi_{32} \equiv [a a b b c d d e e c]$, $\phi_{33} \equiv [a a b b c d e d e c]$. The large $N$ loop splitting vector reads
\begin{equation}
\begingroup
\openup 4pt
    \omega =  \bordermatrix{
        \text{indices} & \cr
        1 &  6 \cr
        2 & 16 \phi_1 \cr
        3 &  8 \phi_1 \cr
        4 &  24 \phi_2 + 6 \phi_1^2 \cr
        5 &  6 \phi_2 + 10 \phi_3 + \frac{10}{3} \phi_1^2 \cr
        6 & 20 \phi_2 + \frac{22}{3} \phi_1^2 \cr
        7 &  4 \phi_2 + 8 \phi_3 + \frac{10}{3} \phi_1^2 \cr
        8 &  12 \phi_2 + 6 \phi_1^2 \cr
        \vdots & \vdots \cr
    } \, .
\endgroup
\end{equation}
The collective potential is
\begin{equation}
    V_{\col}(\phi) = \frac{1}{8} \omega^{\rm{T}} \Omega^{-1} \omega + \frac{g_{\rm{YM}}^2}{2} (\phi_2 - \phi_3) \, ,
\end{equation}
and the virial theorem gives $\omega^{\rm{T}} \Omega^{-1} \omega = 8 g_{\rm{YM}}^2 (\phi_2 - \phi_3)$, so that $E / N^2 = \frac{3}{2} g_{\rm YM}^2 (\phi_2 - \phi_3)$. Besides $\Omega \succeq 0$, traces of pairs of non-singlet words also gives $\mathrm{O}(3)$-invariant loops and hence additional positivity constraints. One of the simplest examples is to take a sequence of operators 
\begin{equation}
    \biggl\{
    O_i \equiv M_a R^i \biggm| 
    R \equiv \sum_{a=1}^{3} M_a^2 \, , \,
    i = 0, 1, 2, \dots
    \biggr\} \, ,
\end{equation}
the Gram matrix built from this sequence is positive semidefinite,
\begin{equation}
    \mathcal{G} \succeq 0 \, , 
    \quad \mathcal{G}_{ij} \equiv \phi(\mathcal{O}_i^\dagger \mathcal{O}_j) = \phi(R^{i+j+1}) \, .
\end{equation}
Here the flavor indices of the matrices are summed over. More generally, the Gram matrix of any set of words carrying a free flavor index, contracted over that index, is positive semidefinite, and gives constraints beyond the symmetry reduced loop joining matrix $\Omega \succeq 0$. 

\begin{table}[!thb]
\begin{center}
\caption{Ground state energy and the three lowest invariant loops of the three-matrix Yang--Mills model at $g_{\rm{YM}}=1$ and truncation $\ell_{\max} = 7$ ($L=12$), with cages $\langle \phi(R^{n+1}) \rangle < S \langle \phi(R^{n}) \rangle$ for several~$S$. The last row shows the published master field fits \cite{Rodrigues:2025sbu}.} 
\label{tab:cage-comparison-g-paper-one}
\begin{tabular}{rcccc}
\toprule $S$ & $E_{\mathrm{g}}$ & $[aa]$ & $[aabb]$ & $[abab]$ \\
\midrule
3 & 1.36723070 & 1.31595313 & 2.24536442 & 1.33387728 \\
4 & 1.36712139 & 1.31653855 & 2.24885736 & 1.33744311 \\
8 & 1.36705636 & 1.31683209 & 2.25083525 & 1.33946434 \\
16 & 1.34970241 & 1.44302344 & 2.53985597 & 1.64005436 \\
\midrule MF & 1.36752999 & 1.31500770 & 2.24004720 & 1.32836100 \\
\bottomrule
\end{tabular}
\end{center}
\end{table}

\begin{figure}[!thb]
\begin{center}
    \includegraphics[width=0.6\textwidth]{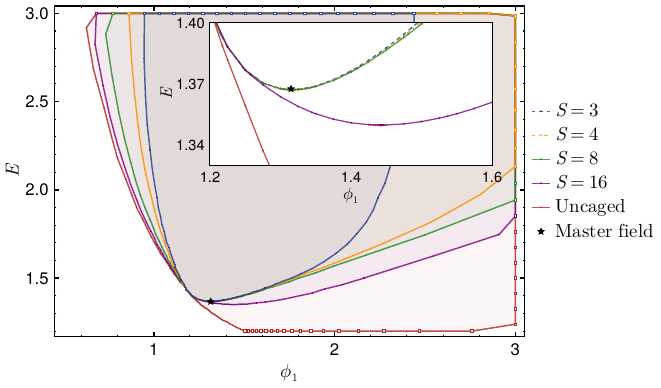}
    \caption{Feasible region search in the $(\phi_1, E)$ plane for various cages $S$. The inset plot exhibits a refinement of the boundary lines separating the feasible and infeasible regions.}
    \label{fig:3MQM_cage_effect}
\end{center}
\end{figure}

This model was studied with master field optimization in \cite{Rodrigues:2025sbu}, at loop truncation $\ell_{\max} = 6$ with 9503 loops after removing the cyclic redundancy. We perform the collective bootstrap at $\ell_{\max} = 7$ for comparison, so that there are 305 loop variables and 6 lifted $Y$ variables.
The constraint due to the Gram matrix $\mathcal{G} \succeq 0$ bounds the moments from below but not from above. To prevent the high-lying loops from becoming too large in the bootstrap process, we also add the cages $\langle \phi(R^{n+1}) \rangle < S \langle \phi(R^{n}) \rangle $ for a positive number $S$. These artificial constraints are not implied by the dynamics. They are equivalent to assuming that the large $N$ eigenvalue distribution of the operator $R$ in the normalization of the loops is supported below $S$, and they are satisfied by the true ground state only if $S$ exceeds the edge of this support. For such $S$ the bound \eqref{eq:bound_hierarchy} remains valid, whereas a smaller $S$ can exclude the true solution. Under the coupling rescaling, a cage specified at $g_{\rm YM}=1$ transforms as $S(g_{\rm YM})=S(1)g_{\rm YM}^{-2/3}$. 
The effect of the cages is shown in \cref{tab:cage-comparison-g-paper-one} and \cref{fig:3MQM_cage_effect}. Without cages, the search terminates at a point with $\langle \phi_2 \rangle \approx 31$ and $\langle \phi_3 \rangle \approx 30$, an order of magnitude larger than the master field values. This is consistent with \eqref{eq:bound_hierarchy}, but shows that at $\ell_{\max} = 7$ these constraints alone are not strong enough to confine the loops, and lower energies are reached along directions in which the individual loops grow large while their difference $\phi_2 - \phi_3$ remains finite. With the cages, the results become insensitive to $S$. For $S = 3, 4, 8$ the energy varies only at the level of $10^{-4}$. Therefore, moderate tight cages produce a stable energy scale and low loop values moments, shrinking the cage $S$ further could eventually raise the energy substantially or make the constraints infeasible.

\begin{figure}[!t]
\begin{center}
    \includegraphics[width=0.45\textwidth]{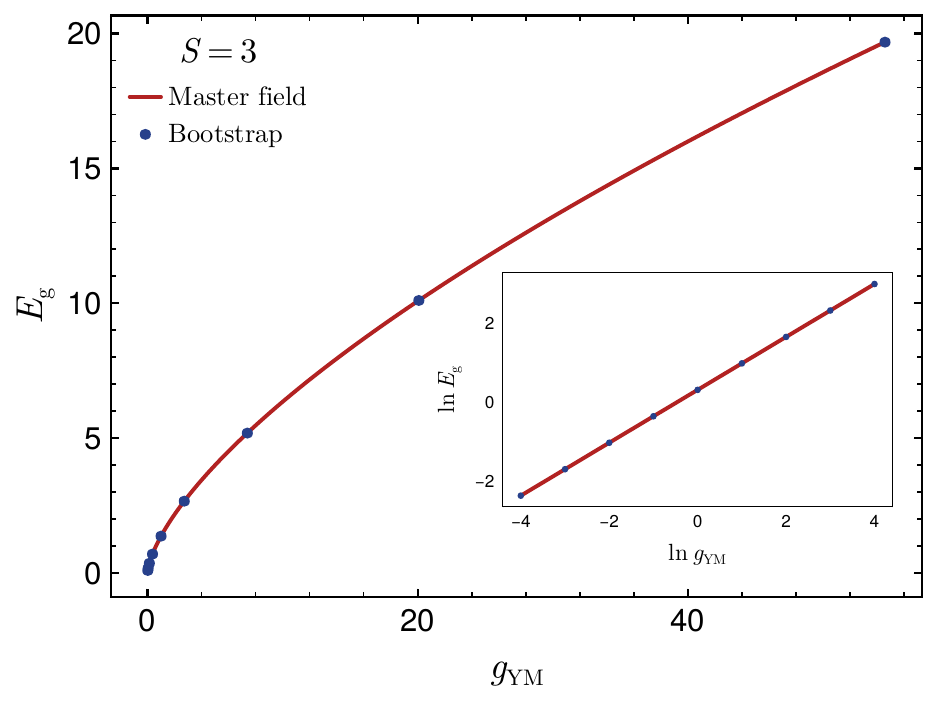}
    \caption{Ground state energy of the three-matrix Yang--Mills model versus $g_{\rm{YM}}$, compared with the master field result of \cite{Rodrigues:2025sbu}.}
    \label{fig:E_3MQM}
\end{center}
\end{figure}

\begin{figure}[!t]
\begin{center}
    \begin{subfigure}[t]{0.45\textwidth}
        \includegraphics[width=0.99\textwidth]{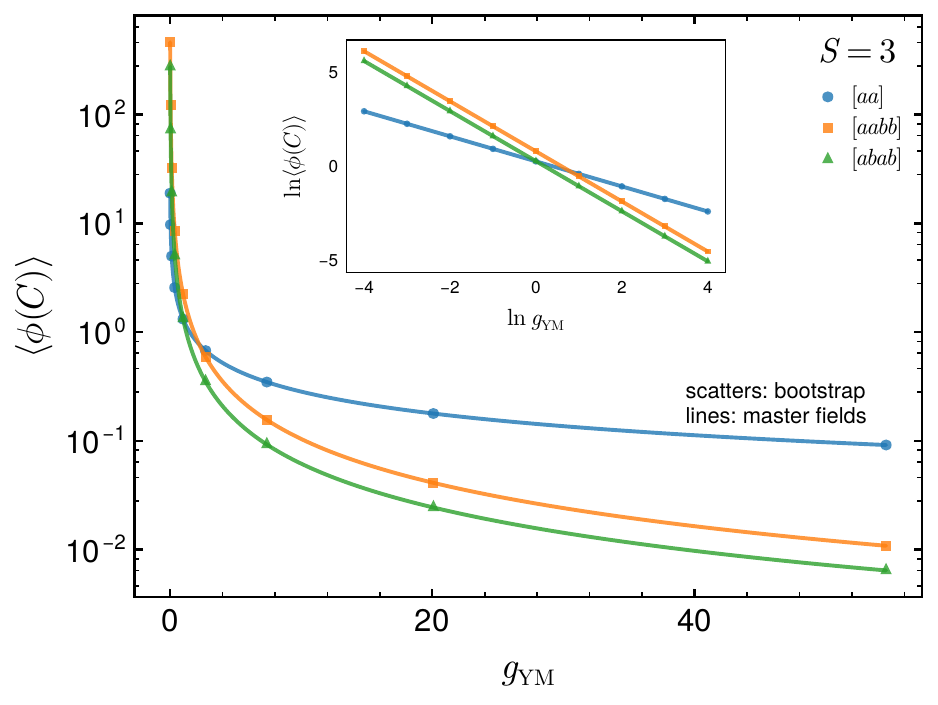}
    \end{subfigure}
    \begin{subfigure}[t]{0.45\textwidth}
        \includegraphics[width=0.99\textwidth]{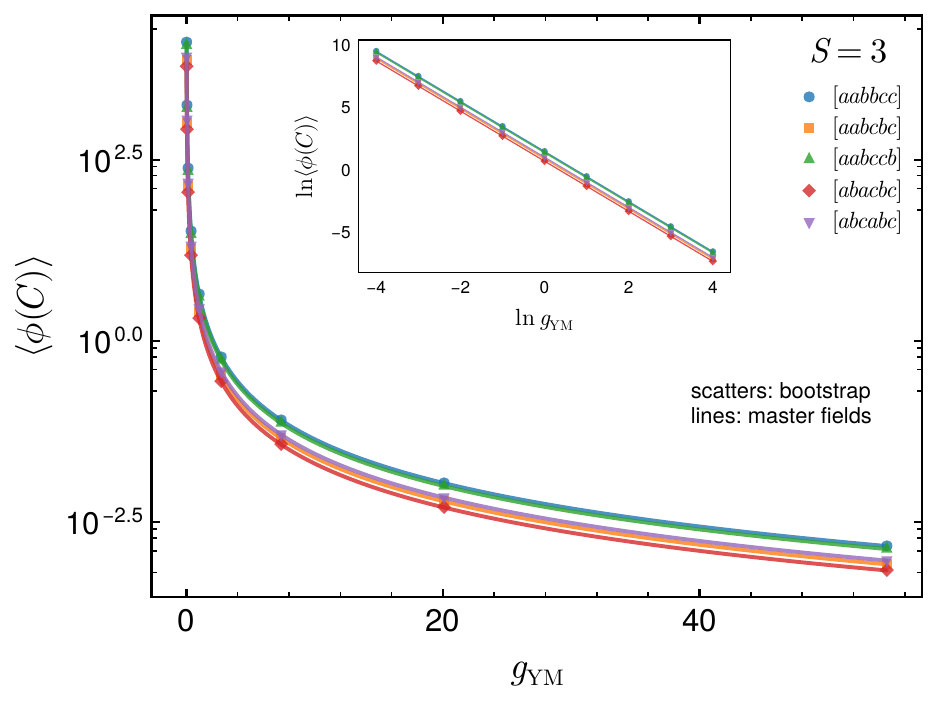}
    \end{subfigure}
    \caption{The lowest loops of the three-matrix Yang--Mills model versus $g_{\rm{YM}}$, loops of length 2 and 4 (left) and of length 6 (right).} 
    \label{fig:loop_3MQM}
\end{center}
\end{figure}

The comparisons of the bootstrap results and master field optimization results are presented in \Cref{fig:E_3MQM,fig:loop_3MQM}, and agreement at the $0.1\%$ level is found for the energy. The fitting from the bootstrap results for $S=3$ gives 
\begin{equation}
    E_{\rm{g}} \approx 1.3672 \, g_{\rm{YM}}^{2/3} \, ,
\end{equation}
which is lower than the master field result $E_{\rm{g}}^{\rm{MF}}=1.36752999\,g_{\rm{YM}}^{2/3}$ \cite{Rodrigues:2025sbu} by $0.022\%$. Fits of the lowest loops give
\begin{align}
    \langle \phi(aa) \rangle & \approx 1.3160 \, g_{\rm{YM}}^{-2/3} \, , \\
    \langle \phi(aabb) \rangle & \approx 2.2454 \, g_{\rm{YM}}^{-4/3} \, , \\
    \langle \phi(abab) \rangle & \approx 1.3339 \, g_{\rm{YM}}^{-4/3} \, .
\end{align}
The corresponding angle
\begin{equation}
    \theta_{ab} \equiv - \frac{N}{2} \frac{\langle \tr([M_a, M_b]^2) \rangle}{\langle \tr(M_a^2) \rangle \langle \tr(M_b^2) \rangle} 
    = \frac{3}{2} \frac{\langle \phi_2 - \phi_3 \rangle}{\langle \phi_1 \rangle^2} 
    \approx 0.7895 \, ,
    \quad
    \forall a \neq b \, ,
\end{equation}
agrees with the master field value $0.791(2)$ \cite{Rodrigues:2025sbu}, and is again independent of the coupling, as in the two-matrix case \cite{Han:2020bkb,Mathaba:2023non}. As in the two-matrix case, the virial relation implies $E_{\rm g} = \theta_{ab} \, g_{\rm YM}^2 \langle \phi_1 \rangle^2$, and the fitted values indeed satisfy this relation.

\section{Discussion}
\label{sec:discussion}

We have developed a collective bootstrap framework for large $N$ matrix quantum mechanics in terms of single trace coordinate moments (loops). Building on the collective field formulation~\cite{Jevicki:1979mb,Jevicki:1980zg,Jevicki:1983hb}, the method organizes the kinetic constraints through loop joining and splitting, so that words involving momentum matrices never appear as independent variables. The inverse joining matrix is traded for a positive semidefinite block using the virial theorem and a Schur complement, and products of loops are lifted to auxiliary variables as in the relaxation bootstrap~\cite{Kazakov:2021lel,Kazakov:2022xuh}. The resulting SDP gives a lower bound on the planar ground state energy, and a sequential SDP based on the log-det heuristic searches for factorized configurations. The one-matrix and decoupled two-matrix models provide analytic benchmarks, while the interacting two- and three-matrix models allow comparisons with master field optimization.

Several aspects of the method deserve further study. First, the rank-one search is a heuristic and does not certify rigorous optimality. Alternative rank-reduction schemes \cite{Recht:2010rank}, or a branch-and-bound treatment of the nonconvex constraint, could narrow this gap. Second, only one dynamical SD equation, i.e.\ the virial theorem, is imposed explicitly, and energy minimization takes the place of the others. The higher dynamical equations derived in \cref{app_sub:one-MQM_SD_hierarchy} could be added after further lifting, and are expected to improve the precision of the loops, which are determined less accurately than the energy. Finally, the number of lifted variables grows quadratically with the number of short loops, so that an efficient use of symmetries, as in~\cite{Lin:2024vvg,Lin:2025srf}, will be essential for larger systems.

Once a factorized (rank-one condition of $Y$) solution is found, the collective Hamiltonian also determines the spectrum of $O(1)$ fluctuations around it, through the joining matrix and the Hessian of the collective potential at the large $N$ background solution \cite{Koch:2021yeb,Mathaba:2023non,Rodrigues:2025sbu}. Computing this spectrum from the bootstrap solution would combine the main strengths of the bootstrap and master field approaches.

An immediate application is the bosonic sector of the nine-matrix BFSS and BMN models~\cite{Banks:1996vh,Berenstein:2002jq}, whose thermal properties have been studied by Monte Carlo \cite{Bergner:2019rca}. Extending the method to the full supersymmetric theories is possible in principle. Existing bootstrap studies of D0-brane quantum mechanics \cite{Lin:2023owt,Lin:2024vvg,Lin:2025srf} and of supersymmetric quantum mechanics and the Marinari--Parisi matrix model \cite{Laliberte:2025xvk} provide concrete benchmarks. Beyond the ground state, the thermal bootstrap \cite{Cho:2024kxn,Cho:2025vws,Klebanov:2026rfc} combines SD equations with thermal inequalities and semidefinite approximations of the matrix logarithm. A thermal collective bootstrap would likewise require thermal constraints beyond the zero-temperature collective potential, such as the KMS condition. In parallel, the doubled collective formulation of matrix thermofield theory~\cite{Jevicki:2025ybt} offers an alternative starting point. It would also be interesting to extend the framework to time-dependent \cite{Lawrence:2024mnj} and nonequilibrium dynamics \cite{Cho:2025nlv}. We hope to address these questions in future work.

\section*{Acknowledgements}

We thank Antal Jevicki and Robert de Mello Koch for illuminating discussions and collaboration at an early stage of this work. GPT6.1 sol was used for the coding of the three-matrix model, and Opus 5.5 was used for manuscript improvement. This work is partially supported by the National Science Foundation of China (NSFC), Individual Grant No.12374138.

\appendix

\section{A simple example: the vector model}
\label{app:vector_model}

In this appendix we present the simplest example of the collective bootstrap, the $\mathrm{O}(N)$ vector model in quantum mechanics, with Hamiltonian
\begin{equation}
    H = \frac{1}{2} \vec{p}^{\, 2} + \frac{1}{2} \vec{x}^{\, 2} + \frac{g}{4 N} (\vec{x}^{\, 2})^2 \, ,
\end{equation}
where $\vec{x} = x^i \mathbf{e}_i$ with $i = 1, \dots, N$. For the collective variable $\phi = \vec{x}^{\, 2} / N$ we have
\begin{equation}
    \Omega = 4 \phi \, , \quad \omega = 2 \, ,
\end{equation}
and the collective Hamiltonian is
\begin{align}
    H_{\col} & = \frac{1}{2 N} \Pi \Omega \Pi + N V_{\col}(\phi) \, , \\
    V_{\col} & = \frac{1}{8} \omega \Omega^{-1} \omega + \frac{1}{2} \phi + \frac{g}{4} \phi^2 \, .
\end{align}
At large $N$, varying $V_{\col}$ with respect to $\phi$ gives the gap equation
\begin{equation}
    1 + g \phi = \frac{1}{4 \phi^2} \, ,
\end{equation}
which resums the bubble diagrams. Since only $\phi > 0$ is required, a planar solution also exists for negative $g$, down to the critical value $g_{\rm{c}} = - 4 /(3 \sqrt{3})$, where $\phi_{\rm{c}} = \sqrt{3}/2$.

We now solve this problem in the bootstrap spirit, i.e.\ by imposing positivity constraints in the space of variables. The first step is to linearize the problem so that SDP applies. The Jacobian term $\omega \Omega^{-1} \omega / 8 = 1/(8 \phi)$ is nonlinear, and we introduce the variable $\mu \equiv \omega \Omega^{-1} \omega$ to linearize it. The potential also contains the ``multi trace'' term $g \phi^2 / 4$, which is quadratic, and we introduce
\begin{equation}
    y \equiv \phi^2
\end{equation}
to linearize it. In other words, we \emph{enlarge} the set of variables from $\{\phi\}$ to $\{\phi, \mu, y\}$. The second step is to eliminate as many variables as possible using exact relations. The virial theorem gives
\begin{equation}
    \mu = 4 (\phi + g y) \, ,
\end{equation}
which eliminates $\mu$ and leaves the variables $\{\phi, y\}$. As a bonus, the energy becomes linear,
\begin{equation}
    E = \phi + \frac{3 g}{4} y \, .
\end{equation}
The last step is to impose the constraints. One is $\Omega = 4 \phi \geq 0$. In addition, we must impose the relation between $\phi$ and $y$. This is done with the lifted matrix $Y = v \otimes v$, $v \equiv (1, \phi)$, which is positive semidefinite and of rank one; replacing $\phi^2$ by $y$ we have
\begin{equation}
    Y = \begin{pmatrix}
        1 & \phi \\
        \phi & y
    \end{pmatrix} \succeq 0 \, ,
    \quad
    \text{and}
    \quad
    \rank(Y) = 1 \, .
\end{equation}
Since equality constraints of this type are hard to impose numerically, we \emph{relax} the definition of $\mu$ to an inequality, such that the consistency condition lies on the boundary of the allowed region. Using the Schur complement,
\begin{equation}
    \begin{pmatrix}
        \Omega & \omega \\
        \omega & \mu
    \end{pmatrix}
    =
    \begin{pmatrix}
        \Omega & \omega \\
        \omega & 4 (\phi + g y)
    \end{pmatrix} \succeq 0 \, .
\end{equation}
Consistency requires the stronger condition $\mu = \omega \Omega^{-1} \omega$, i.e.\ saturation. The Schur complement condition also implies $\Omega \succeq 0$.

\begin{figure}[t!]
\begin{center}
    \includegraphics[width=0.5\textwidth]{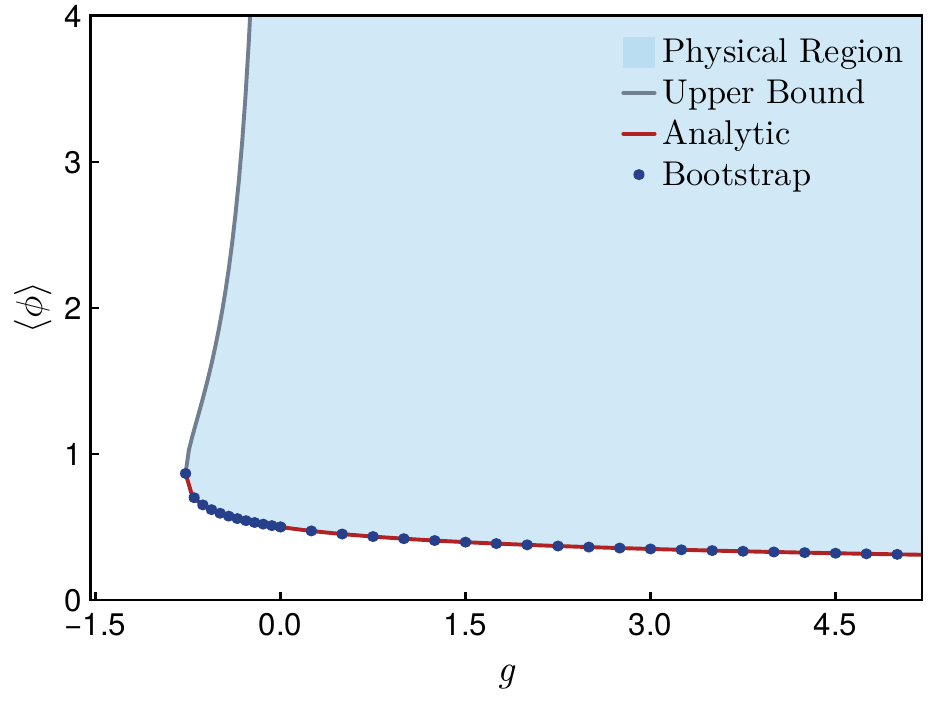}
    \caption{Collective bootstrap of one-dimensional vector model.}
    \label{fig:vector_model}
\end{center}
\end{figure}

\begin{figure}[htb!]
\begin{center}
    \includegraphics[width=0.45\textwidth]{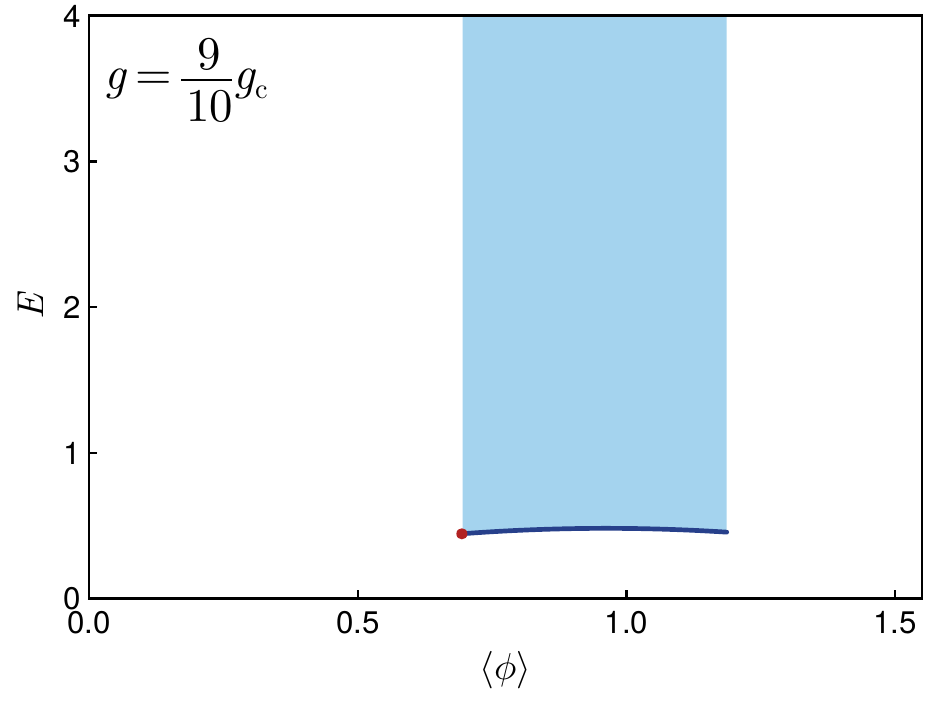}\hfill
    \includegraphics[width=0.45\textwidth]{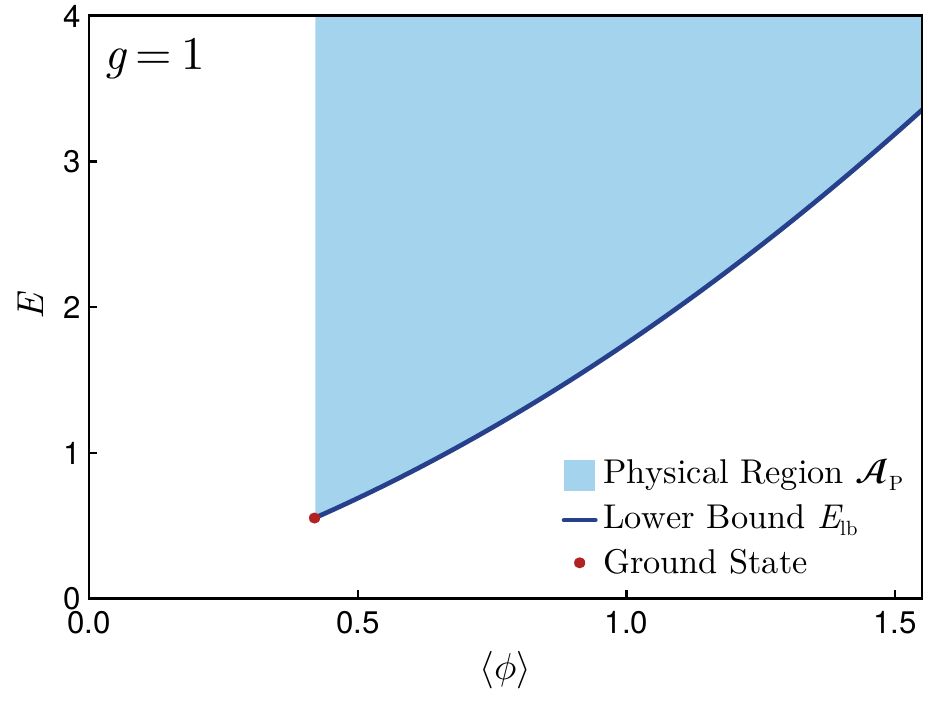}
    \caption{Physically allowed regions for the one-dimensional vector model for fixed $g$. Left: $g = 9 g_{\rm c} / 10$. Right: $g = 1$.}
    \label{fig:vector_model_physical_regions}
\end{center}
\end{figure}

Having linearized the problem with the auxiliary variables $y$ and $\mu$, at the cost of additional constraints relaxed to inequalities, we arrive at the SDP
\begin{align}
    \text{minimize } &  
    E + \tr(Y) = \phi + \frac{3 g}{4} y + 1 + y \, , \nonumber \\
    \text{subject to } & 
    \begin{pmatrix}
        1 & \phi \\
        \phi & y
    \end{pmatrix} \succeq 0
    \text{ and }
    \begin{pmatrix}
        \Omega & \omega \\
        \omega & 4 (\phi + g y)
    \end{pmatrix} \succeq 0 \, ,
\end{align}
where we have added $\tr(Y)$ to the objective function to favor low-rank solutions. This simple SDP is easily implemented, e.g.\ with \texttt{SemidefiniteOptimization} in \texttt{Mathematica}. \Cref{fig:vector_model} shows the resulting $\phi$ versus $g$, in precise agreement with the gap equation, and the condition $\rank(Y)=1$ is easily reached by the solver. Scanning $g$ also locates the critical coupling, beyond which no feasible point exists. \Cref{fig:vector_model_physical_regions} shows the feasible regions for $g = 9 g_{\rm c} / 10$ and $g = 1$.

\section{Analytic bootstrap for one-matrix quantum mechanics}
\label{app:analytic_bootstrap_one_MQM}

\subsection{Recursion relations}

\begin{figure}[t!]
\begin{center}
    \includegraphics[width=0.45\textwidth]{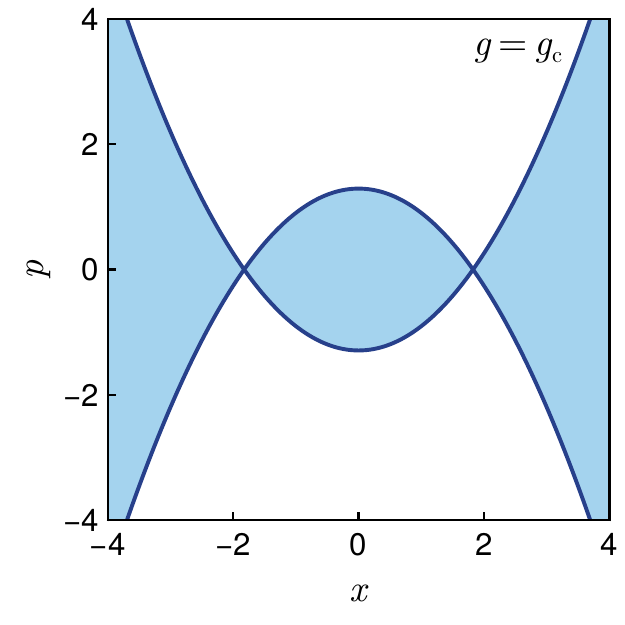} \hfill
    \includegraphics[width=0.45\textwidth]{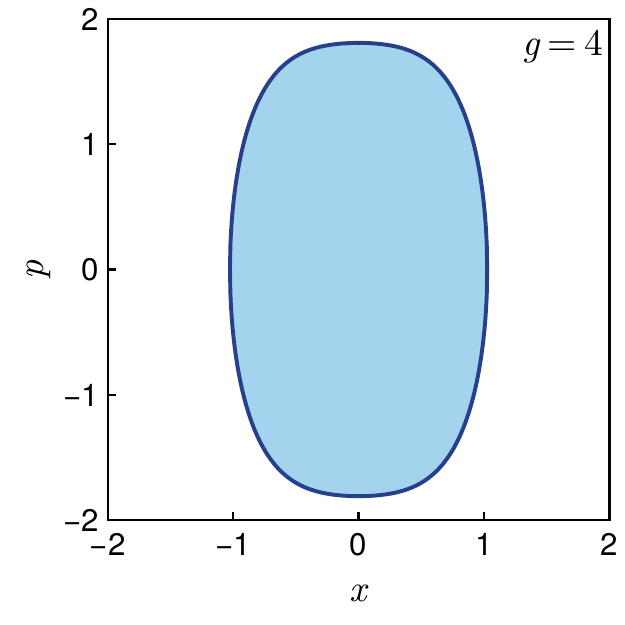}
    \caption{The distribution function $u(x, p)$ in phase space. Left: $g = g_{\rm c}$. Right: $g = 4$.}
    \label{fig:u_plot}
\end{center}
\end{figure}

In this appendix we derive recursion relations for the loops of the one-matrix model \eqref{eq:one-MQM_Hamiltonian}. At large $N$ the singlet sector is equivalent to free fermions \cite{Brezin:1977sv}, and the ground state is described by a Fermi sea in phase space with occupation function
\begin{equation}
    u(x, p) = \theta(\epsilon_{\rm F} - h(x, p)) \, , \quad   h(x, p) = \frac{p^2}{2} + v(x) \, ,
\end{equation}
where $v(x) = x^2 / 2 + g x^4 / 4$ is the potential and $\epsilon_{\rm F} = v(\Lambda)$ is the Fermi energy. Integrating $u$ over $p$ gives the eigenvalue density
\begin{equation}
    \rho(x) = \int \frac{\dd{p}}{2\pi} u(x, p)  
    = \int_{-p_{\rm F}}^{p_{\rm F}} \frac{\dd p}{2\pi} = \frac{1}{\pi} p_{\rm F}\, ,
\end{equation}
where $p_{\rm F} = \sqrt{2 (\epsilon_{\rm F} - v(x))}$ is the Fermi momentum. The Fermi sea is illustrated in \cref{fig:u_plot}.

\begin{figure}[!thb]
\begin{center}
    \begin{subfigure}[t]{0.45\textwidth}
        \includegraphics[width=1\textwidth]{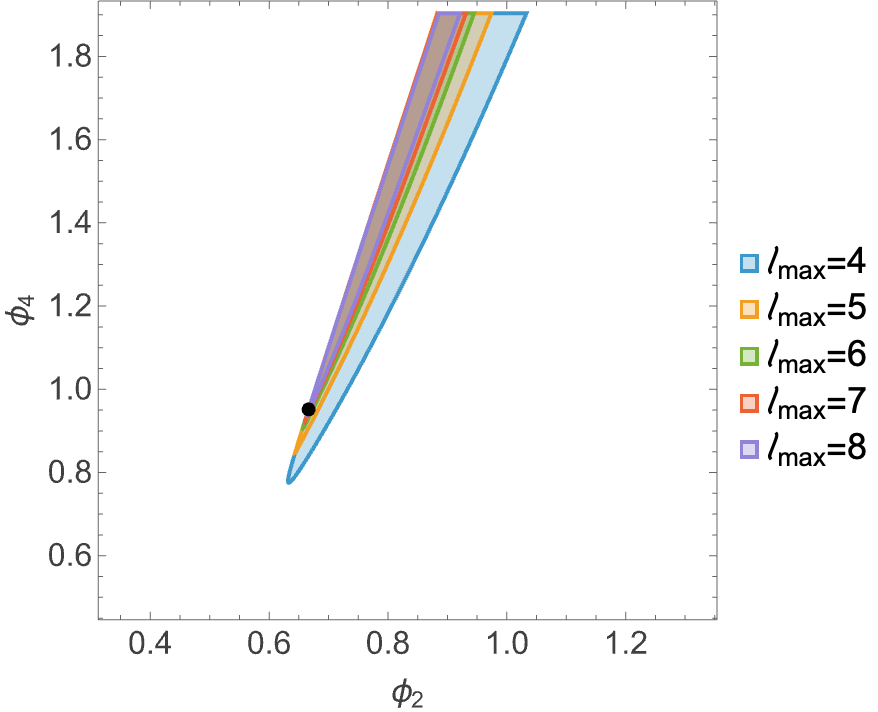}
        \caption{$g = g_{\rm{c}}$.}
    \end{subfigure}
    \begin{subfigure}[t]{0.45\textwidth}
        \includegraphics[width=1\textwidth]{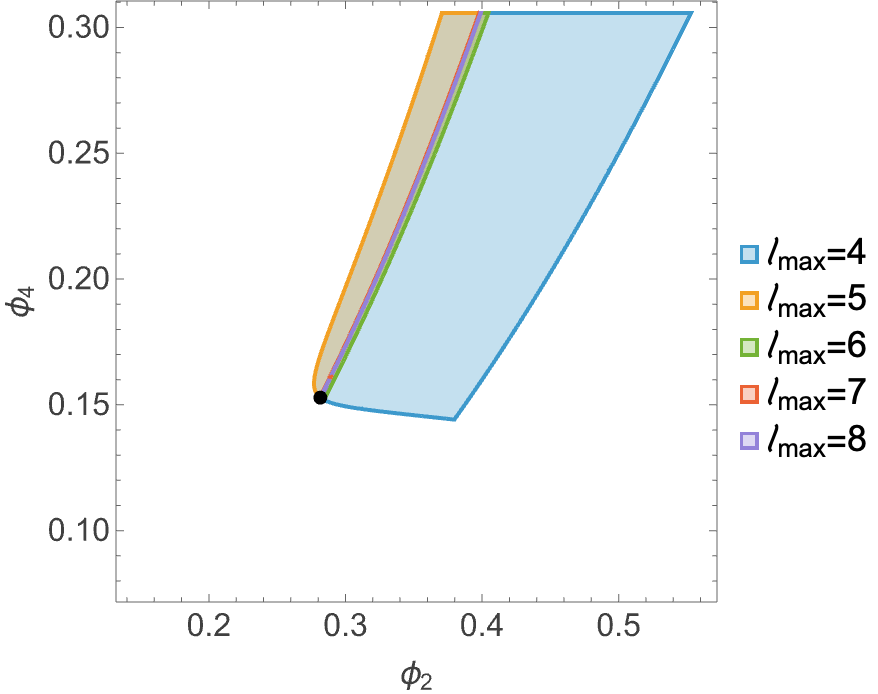}
        \caption{$g = 4$.}
    \end{subfigure}
    \caption{Feasible regions with recursion relations \eqref{eq:loop_eq_analytic} at different truncation lengths. The black points denote the ground states.}
    \label{fig:1MQM_recursion_regions}
\end{center}
\end{figure}

In this description the SD equation $\langle{[H, \mathcal{O}]}\rangle = 0$ becomes the phase space condition
\begin{equation}
    \int \frac{\dd{x} \dd{p}}{2\pi} u(x, p) \{h, \mathcal{O}(x, p)\} = 0 \, ,
\end{equation}
where $\{\cdot, \cdot\}$ denotes the Poisson bracket. For $\mathcal{O} = \tr(M^n P)$ we have $\mathcal{O}(x, p) = x^n p$, and therefore
\begin{equation}
    0 = \int \frac{\dd{x} \dd{p}}{2\pi} u(x, p) [n x^{n-1} p^2 - x^{n}(x + g x^3)] \, .
\end{equation}
The second term in the bracket gives $\phi_{n+1} + g \phi_{n+3}$, while the first term becomes
\begin{align*}
    n \int \frac{\dd x \dd p}{2 \pi} u(x, p) x^{n-1} p^2 & = n \int \dd{x} x^{n-1} \int_{- p_{\rm F}}^{p_{\rm F}} \frac{\dd p}{2\pi} p^2 \\
    & = \frac{n}{3 \pi} \int \dd{x} x^{n-1} \, p_{\rm F}^{3} \\
    & = \frac{n}{3} \int \dd{x} x^{n-1} \, p_{\rm F}^2 \, \rho(x) \\
    & = \frac{n}{3} \int \dd{x} x^{n-1} (2\epsilon_{\rm F} - x^2 - g x^4 / 2) \rho(x) \\
    & = \frac{n}{3} \Bigl(2 \epsilon_{\rm F} \phi_{n-1} - \phi_{n+1} - \frac{g}{2} \phi_{n+3} \Bigr) \, .
\end{align*}
Collecting terms, we obtain
\begin{equation} \label{eq:loop_eq_analytic}
    \underbrace{\frac{n}{3}  \Bigl(2 \epsilon_{\rm F} \phi_{n-1} - \phi_{n+1} - \frac{g}{2} \phi_{n+3} \Bigr)}_{\langle [T, \tr(M^n P)] \rangle} = \underbrace{
    {\color{white}{\frac{1}{1}}\!\!\!\!}
    \phi_{n+1} + g \phi_{n+3}}_{\langle [V, \tr(M^n P)] \rangle } \, , \quad
    n = 1, 2, 3, \dots \, .
\end{equation}
For $n = 1$ this is the virial theorem, which in this form determines the Fermi energy,
\begin{equation}
    \epsilon_{\rm F} = 2 \phi_2 + \frac{7}{4} g \phi_4 \, .
\end{equation}
These equations form a complete set of recursion relations for the ground state loops, equivalent to the SD equations \eqref{eq:boostrap_SD}: given $\phi_2$ and $\phi_4$, all higher loops are determined. One can verify that the analytic solution satisfies them. The feasible regions obtained by combining these recursion relations with positivity are shown in \cref{fig:1MQM_recursion_regions}.

For $\mathcal{O} = \tr(M^n)$, i.e.\ $\mathcal{O}(x, p) = x^n$, the constraints are satisfied identically, since $u$ is even in $p$.

\subsection{Schwinger--Dyson hierarchy}
\label{app_sub:one-MQM_SD_hierarchy}

We now consider the SD equations $\langle [H, \mathcal{O}] \rangle = 0$ \cite{Han:2020bkb} used in the phase space bootstrap of the one-matrix model, and derive their representation in loop space using collective field theory. We focus on the Hermitian one-matrix model; the multi-matrix case proceeds in a similar manner. The single trace operators are traces of the word $\mathcal{C}$ built from the alphabet $\{M, P\}$:
\begin{equation}
    \phi(\mathcal{C}) = \frac{1}{N^{\frac{\len(\mathcal{C})}{2} + 1}} \tr(\mathcal{C}) \, , 
    \quad
    \mathcal{C} = M^{m_1} P^{n_1} M^{m_2} P^{n_2} \dots 
\end{equation}
Each such operator gives a constraint $\langle [H, \phi(\mathcal{C})] \rangle = 0$, which is kinetic if $P \notin \mathcal{C}$ and dynamical otherwise. We consider the two families with the fewest momenta,
\begin{align}
    \phi_{n} \equiv \phi(M^n) \, , \quad 
    \phi_{n,1} \equiv \phi(M^n P) \, . \quad
\end{align}
The first family gives all kinetic constraints, which are independent of the potential, while the second gives dynamical constraints.

Using collective field theory, we express all these constraints in terms of the $\phi_{n}$, obtaining the analog of the loop equations of matrix integrals \cite{Makeenko:1979pb}. It is convenient to introduce
\begin{equation} \label{eq:one-MQM_Q_def}
    Q_n \equiv (\Omega^{-1} \omega)_n = N^{-2} \pdv{\log J}{\phi_n} \, ,
\end{equation}
where $J$ is the Jacobian of the change of variables, and $\Omega$ and $\omega$ are given in \cref{eq:one-MQM_loop_joining,eq:one-MQM_loop_splitting}.

We first consider the kinetic constraints, obtained by commuting $H$ with $\phi_{n} = \tr(M^n) / N^{n/2 + 1}$,
\begin{equation}
    0 = \langle [H, \phi_n] \rangle = \langle [T, \phi_n] \rangle \, .
\end{equation}
Using $[AB, C] = A[B, C] + [A, C]B$, we have
\begin{equation}
    [P^{a}{}_{b}, (M^n)^{c}{}_{d}] = - \ii \sum_{k=0}^{n-1} (M^k)^{c}{}_{b} (M^{n-k-1})^{a}{}_{d} \, ,
\end{equation}
and therefore
\begin{equation}
    0 = \langle [T, \phi_{n}] \rangle = - \frac{\ii n}{2} ( \langle \phi_{n-1,1}^{\dagger} \rangle + \langle \phi_{n-1, 1} \rangle ) \, , 
\end{equation}
where $\phi_{n-1,1}^{\dagger} = \tr(P M^{n-1}) / N^{n/2 + 1}$ is the Hermitian conjugate of $\phi_{n-1, 1}$. This gives the constraint
\begin{equation}
    \langle \tr(P M^{n}) \rangle = - \langle \tr(M^{n} P) \rangle \, .
\end{equation}
On the other hand, moving $P$ in $\tr(M^n P)$ to the left gives the operator identity
\begin{equation}
    \tr(M^n P) = \ii \sum_{k=0}^{n-1} \tr(M^{k}) \tr(M^{n-k-1}) + \tr(P M^n) \, .
\end{equation}
These two equations together give
\begin{equation}
    2 \langle \phi_{n,1} \rangle = \ii \sum_{k=0}^{n-1} \phi_{k} \phi_{n-k-1} \, .
\end{equation}
In collective field theory, this reads
\begin{equation}
    \sum_{k=0}^{\infty} k \phi_{k+n-1} Q_{k} = \sum_{k=0}^{n-1} \phi_{k} \phi_{n-k-1} \, , \quad \forall \, n \geq 1 \, .
\end{equation}
Setting $n = m-1$ and multiplying by $m$, this equation becomes $\Omega Q = \omega$, the defining equation of $Q$ in \eqref{eq:one-MQM_Q_def}. The kinetic constraints are thus built into collective field theory, which is its main advantage for the bootstrap.

Next we consider the dynamical constraints from $\phi_{n,1}$. Using $\langle \tr(P M^n) \rangle = -\langle \tr(M^n P) \rangle$, we obtain
\begin{equation}
    \frac{1}{2} \sum_{k=0}^{n-1} \Bigl( 
        \tr(P M^{n-k-1} P M^{k}) + \tr(M^{n-k-1} P M^{k} P)   
    \Bigr)
    =
    \langle \tr(M^{n} V'(M)) \rangle \, .
\end{equation}
For even $n$ this constraint is trivial by the $\mathbb{Z}_2$ symmetry $M \rightarrow - M$, so we focus on odd $n$. For $n = 1$ it is the virial theorem
\begin{equation}
    2 \langle T \rangle = \langle \tr(M V'(M)) \rangle \, ,
\end{equation}
which in collective field theory translates to
\begin{equation}
    \frac{1}{4} \omega^{\operatorname{T}} \Omega^{-1} \omega = \phi_2 + g \phi_4 \, .
\end{equation}
To express these constraints in loop space, note that the left hand side involves $\tr(P M^{k} P M^{n-k-1})$ and its symmetrizations. This motivates the shifted loop joining
\begin{equation}
    \Omega^{(s)}_{m n} \equiv N^{2-\frac{s}{2}} \tr(M^{s} \pdv{\phi_m}{M} \pdv{\phi_n}{M}) = m \, n \, \phi_{m + n + s - 2} \, .
\end{equation}
Introducing the discrete convolution $(A * B)_s = \sum_{k=0}^{s} A_{s-k} B_{k}$, we can write the constraints as
\begin{equation} \label{eq:loop_eq_bootstrap}
    \frac{n}{4} Q^{\rm{T}} \Omega^{(n-1)} Q - \frac{1}{2} (\phi * \omega)_{n-1} = \phi_{n + 1} + g \phi_{n + 3} \, ,
    \quad
    n = 1, 3, 5, \dots \, ,
\end{equation}
where $Q \equiv \Omega^{-1} \omega$. The second term on the left hand side is a sum of triple products,
\begin{equation}
    \frac{1}{2}  (\phi * \omega)_{n-1}  = \frac{n}{3} \sum_{\substack{a,b,c = 0 \\ a+b+c = n-3}} \phi_a \phi_b \phi_c \, .
\end{equation}
By a direct calculation, one finds that this equation \eqref{eq:loop_eq_bootstrap} is precisely the loop space saddle point equation $\Omega \partial V_{\col} / \partial \phi = 0$.
In the previous subsection, the left hand side of \eqref{eq:loop_eq_bootstrap} was reduced to the simple form \eqref{eq:loop_eq_analytic}, giving loop equations for the one-matrix model. Imposing these equations in the bootstrap could increase the precision, and it would be interesting to derive their analogs for multi-matrix models.

Finally, consider dynamical constraints with more momenta, e.g.\ $\mathcal{O} = \tr(M^m P^n)$ with $n > 1$. The commutator with the kinetic term produces $\tr(M^{m-1} P^{n+1})$, and the commutator with the potential produces $\tr(M^{m+1} P^{n-1})$ and $\tr(M^{m+3} P^{n-1})$. These relations determine the loops with $n+1$ momenta from those with $n-1$. Since from above we can determine $\tr(M^n P)$, then using these constraints we can determine $\tr(M^n P^3)$, $\tr(M^n P^5)$, etc. Therefore, these constraints are redundant.

In summary, for the one-matrix model all kinetic constraints are built into the collective field representation, while the dynamical constraints are equivalent to the saddle point equation $\Omega \partial V_{\col} / \partial \phi = 0$. In this sense the collective formulation is equivalent to the phase space bootstrap. 

\section{Ray search for feasible regions}
\label{app:ray_search}

We summarize the ray bisection method used to map out the feasible regions in \cref{fig:one-MQM_feasible_regions} \cite{El-Showk:2012cjh}. Suppose we would like to find the feasible region in a loop subspace $(x, y)$, say $x \equiv \langle \phi(M^2) \rangle$ and $y \equiv \langle \phi(M^4)\rangle$. In the first step one performs an initialization to find a feasible point $(x_0, y_0)$ that satisfies the constraints. Then we perform a ray search for the boundary of the feasible region at some direction:
\begin{equation}
    (x, y) = (x_0, y_0) + r (\cos\theta, \sin\theta) \, .
\end{equation}
For $r = 0$ this is the feasible point from the initialization. Setting $r_{\min} = 0$ and imposing a cutoff $r_{\max}$, we perform a feasibility check at the middle point with $r_{\rm{mid}} = r_{\max} / 2$, whereby we fix loop variables $\langle\phi(M^2)\rangle = x_0 + r_{\rm{mid}} \cos\theta$ and $\langle\phi(M^4)\rangle = y_0 + r_{\rm{mid}} \sin\theta$. Note that in our case we should also fix the corresponding lifted $Y$ variables. If this point passes the check, then we update $r_{\min} = r_{\rm{mid}}$, and increase the radius $r_{\rm{mid}} \rightarrow (r_{\rm{min}} + r_{\max}) / 2$. Otherwise, we update $r_{\max} = r_{\rm{mid}}$, and decrease the radius $r_{\rm{mid}} \rightarrow (r_{\rm{min}} + r_{\max}) / 2$. Iterating this process until the difference $r_{\max} - r_{\min}$ is smaller than the tolerance, we obtain the boundary of the feasible region along that direction. Performing this binary search and scanning over all $\theta \in [0, 2\pi)$ gives the boundary of the physically allowed region as seen from $(x_0, y_0)$. Since the rank-one condition makes the region nonconvex, this reconstructs the region exactly only if it is star-shaped with respect to $(x_0, y_0)$.

\phantomsection
\addcontentsline{toc}{section}{References}
\bibliographystyle{jhep}
\bibliography{references}

\end{document}